\documentclass{aa}  
\usepackage{graphicx}
\usepackage{color}
\usepackage{hyperref}
\usepackage{txfonts}
\begin{document}

\newcommand{\Ha}{$\rm{H}\alpha$}
\newcommand{\Hb}{$\rm{H}\beta$}
\newcommand{\logU}{$\log(\bar U)$}
\newcommand{\Kdist}{$K_{dist}$}
\newcommand{\EWa}{$EW_\alpha$}
\newcommand{\EWb}{$EW_\beta$}
\newcommand{\fOB}{f(OB)}
\newcommand{\QHHe}{Q$_{0/1}$}
\newcommand{\HII}{\ion{H}{ii}}
\newcommand{\temp}{T$_{\rm e}$}

% FFRO commands

\newcommand{\kms}{km\,s$^{-1}$} 
\newcommand{\hi}{\ion{H}{i}}
\newcommand{\hh}{\ion{H}{ii}}
\newcommand{\nii}{[\ion{N}{ii}]}
\newcommand{\sii}{[\ion{S}{ii}]}
\newcommand{\siii}{[\ion{S}{iii}]}
\newcommand{\oi}{[\ion{O}{i}]}
\newcommand{\oii}{[\ion{O}{ii}]}
\newcommand{\oiii}{[\ion{O}{iii}]}
\newcommand{\ha}{H$\alpha$} 
\newcommand{\hb}{H$\beta$} 
\newcommand{\flux}{erg\,s$^{-1}$\,cm$^{-2}$}

   \title{Photoionization models of the CALIFA HII regions}
   \subtitle{I. Hybrid models}

   \author{C. Morisset \inst{\ref{inst:unam}, \ref{inst:emailchris}}
          \and
          G. Delgado-Inglada\inst{\ref{inst:unam}}
          \and 
          S. F. S\'anchez\inst{\ref{inst:unam}}
          \and 
          L. Galbany\inst{\ref{inst:millenium}, \ref{inst:udc}}
          \and
	      R. Garc\'\i a-Benito\inst{\ref{inst:iaa}}
          \and 
          B. Husemann\inst{\ref{inst:eso}}
          \and 
          R. A. Marino\inst{\ref{inst:switz}, \ref{inst:madrid}}
          \and
          D. Mast\inst{\ref{inst:iaa}}         
          \and 
          M. M. Roth\inst{\ref{inst:leibniz}}
          \and 
          CALIFA collaboration\inst{\ref{inst:califa}}
          }

   \institute{Instituto de Astronomía, Universidad Nacional Autonoma de Mexico, Apdo. Postal 70264, 04510, Mexico D.F., Mexico 
           \label{inst:unam}
              \and 
              Millennium Institute of Astrophysics, Chile
           \label{inst:millenium}
			  \and 
              Departamento de Astronomía, Universidad de Chile, Camino El Observatorio 1515, Las Condes, Santiago, Chile
           \label{inst:udc}
              \and
              Instituto de Astrof\'{\i}sica de Andaluc\'{\i}a (CSIC), Glorieta de la Astronom\'\i a s/n, 
              Aptdo. 3004, E18080-Granada, Spain 
           \label{inst:iaa}
              \and 
              European Southern Observatory (ESO), Karl-Schwarzschild-Str.2, D-85748 Garching b. München, Germany
           \label{inst:eso}
              \and 
              Department of Physics, Institute for Astronomy, ETH Zürich, CH-8093 Zürich, Switzerland
              \label{inst:switz}
              \and
              Departamento Astrofísica, Universidad Complutense de Madrid, Avda.Ciudad Universitaria s/n 28040 - Madrid, Spain
           \label{inst:madrid}
              \and 
              Leibniz Institut für Astrophysik, An der Sternwarte 16, 14482 - Potsdam, Germany 	 		              		 
           \label{inst:leibniz}
              \and 
              \url{http://califa.caha.es/}
           \label{inst:califa}
              \and
			  e-mail: \href{mailto:chris.morisset@gmail.com}{chris.morisset@gmail.com}\\
              ORCID 0000-0001-5801-6724
           \label{inst:emailchris}
             }

   \date{Received 04 2016; accepted 06 2016}
 
   \abstract{
      Photoionization models of \ion{H}{ii} regions require as input a description of the ionizing SED (Spectral Energy Distribution) and of the gas distribution, in terms of ionization parameter $U$ and chemical abundances (e.g. O/H and N/O). A strong degeneracy exists between the hardness of the SED and $U$, which in turn leads to high uncertainties in the determination of the other parameters, including abundances. One way to resolve the degeneracy is to fix one of the parameters using additional information.
      
For each of the $\sim$ 20,000 sources of the CALIFA \ion{H}{ii} regions catalog, a grid of photoionization models is computed assuming the ionizing SED being described by the underlying stellar population obtained from spectral synthesis modeling. The ionizing SED is then defined as the sum of various stellar bursts of different ages and metallicities. This solves the degeneracy between the shape of the ionizing SED and $U$.
The nebular metallicity (associated to O/H) is defined using the classical strong line method O3N2 (which gives to our models the status of "hybrids"). The remaining free parameters are the abundance ratio N/O and the ionization parameter $U$, which are determined by looking for the model fitting [\ion{N}{ii}]/H$\alpha$ and [\ion{O}{iii}]/H$\beta$. The models are also selected to fit [\ion{O}{ii}]/H$\beta$. This process leads to a set of $\sim$ 3,200 models that reproduce simultaneously the three observations.

We find that the regions associated to young stellar bursts (i.e., ionized by OB stars) suffer leaking of the ionizing photons, the proportion of escaping photons having a median of 80\%. The set of photoionization models satisfactorily reproduces the electron temperature derived from the [\ion{O}{iii}]$\lambda$4363/5007 line ratio. We determine new relations between the nebular parameters, like the ionization parameter $U$ and the [\ion{O}{ii}]/[\ion{O}{iii}] or [\ion{S}{ii}]/[\ion{S}{iii}] line ratios. A new relation between N/O and O/H is obtained, mostly compatible with previous empirical determinations (and not with previous results obtained using photoionization models). A new relation between $U$ and O/H is also determined.

All the models are publicly available on the Mexican Million Models database 3MdB.
}

   \keywords{Interstellar medium (ISM), ISM: abundances, HII regions, Galaxies: ISM}

   \maketitle
%
%________________________________________________________________

\section{Introduction}

Classical \HII\ regions are large, low-density clouds of partially ionized gas in which star formation has recently taken place ($<$15 Myr). The short-lived blue stars forged in these regions emit large amounts of ultraviolet radiation that ionizes the surrounding gas. They span a wide range of physical scales, from a few parsecs, like the Orion nebula (D$\sim$8~pc), or even smaller \citep{ander14}, to hundreds of parsecs, such as 30~Doradus (D$\sim$20~pc), NGC\,604 (D$\sim$460~pc), or NGC\,5471 (D$\sim$1~kpc) as reported by \cite{oey03} and \cite{rgb2011}. These latter ones are the prototypes of the extragalactic giant \HII\ regions found frequently in the disks of spiral galaxies \citep[e.g.][]{hodg83,dott87,dott89,knap98}, or starburst and blue compact galaxies \citep[e.g.][]{kehrig08,angel09,cairos12}.

\cite{1981Baldwin_pasp93} first proposed the \oiii$\lambda$5007/\Hb\ versus \nii$\lambda$6584/\Ha\ diagnostic diagram (now known as the BPT diagram) to separate emission-line objects according to the main gas excitation mechanism: normal \HII\ regions, planetary nebulae, and objects photoionized by a harder radiation field. The latter can be produced by either a power-law continuum from an AGN (Active Galactic Nuclei), shock excitation, planetary nebulae central stars or even post-AGB (Asymptotic Giant Branch) stars \citep[e.g.][]{binn94,2008Stasinska_mnra391, binn09,mori09,2009Flores-Fajardo_Revi45,kehrig12,sign13,2013Papaderos_aap555}.
\citet{1987Veilleux_apjs63} and \citet{osterbrock89} extended and refined this classification scheme, incorporating new diagnostic diagrams. \cite{osterbrock89} used theoretical photoionization models to infer the demarcation line between star-forming (SF) and AGN galaxies, and added two new diagnostics diagrams that exploit the \oi/\Hb\ versus \sii/\Ha\ line ratios. \cite{dopita00} and \cite{kewley01} combined stellar population synthesis and photoionization models to build the first purely theoretical classification scheme for separating pure AGN from galaxies hosting star formation, and \cite{2003Kauffmann_mnra346} used SDSS \citep[Sloan Digital Sky Survey][]{2000York_aj120} data to observationally constrain these classifications.

In essence, these models assume that the main factors that control the emission line spectrum are the chemical abundances of the heavy elements in the gas phase within an \HII\ region (oxygen being the most important), the shape of the ionizing radiation spectrum, and the geometrical distribution of gas with respect to the ionizing sources. Generally speaking, all the geometrical factors are subsumed into a single factor, the ionization parameter $q$ (with dimensions $cm\ s^{-1}$) or the (dimensionless) ionization parameter $U = q/c$. They also assume  {\tt a priori} that these parameters are independent, and thus these models are presented as grids of oxygen abundance, ionization parameter, shape of the ionizing spectrum (effective temperature or stellar burst age) and sometimes electron densities.

Most of the present day knowledge of these regions is based on the comparison of the predictions between these photoionization models and the largest accessible databases for the observed properties of \HII\ regions. However, in many cases, the samples/catalogs are limited in number (a few hundred of \HII\ regions), and/or biased (\HII\ hosted by Sc/Sd galaxies, due to the better contrast with the continuum). This has been recently overcome by the advent of large IFU surveys that have provided large catalogs of \HII\ regions/aggregations with spectroscopic information (of the order of thousands), over unbiased sample of galaxies \citep[from E to Sds, see e.g.][]{2016Marino_aap585, 2016Sanchez-Menguiano_aap587}.

This is the case of the CALIFA (Calar Alto Legacy Integral Field spectroscopy Area survey) survey \citep{sanchez12b}, that has acquired IFU (Integral Field Unit) data of a sample of $\sim$600 galaxies in the Local Universe ($0.005<z<0.03$), covering the full optical extension of these galaxies \citep[see][for more information on the sample]{2014Walcher_aap569}. This survey has created one of the largest catalogs of \HII\ regions/aggregations, with more than 20,000 ionized regions, with spectroscopic information covering most of the typical emission lines in the optical wavelength range from \oii$\lambda$3727 to \sii$\lambda$6731, and with an accurate spectral modeling and subtraction of the underlying stellar population.

One of the main problems in the determination of the properties of HII regions is that some of the parameters that describe these properties act on very similar ways on the observations. This is the case for the softness of the ionizing radiation and the ionization parameter $U$. Both change the ionization state of the nebula, in particular the line ratios involving two subsequent ions (e.g. \oii/\oiii). The best way to resolve this degeneracy is to find a method to determine one of the two parameters using an alternative observable. Combining the output from spectral synthesis modeling of the CALIFA observations gives us access to the softness of the ionization field, kipping only $U$ to be determined.

In this article we use this extensive catalog to create an 'ad-hoc' grid of photoinization models, with the properties of the ionizing sources {\it a priori} provided by the analysis of the stellar populations, in order to understand the physical conditions of these nebulae. We also use an {\it a priori} determination of the O/H metallicity indicator (using a strong line method) to only have the ionization parameter $U$ and the N/O abundance ratio as free parameter. We are actually doing a work similar to \citet{2010Perez-Montero_mnra404} but using a much larger set of observations.

The paper is organized as follows: the Sec.~\ref{sec:califa} describes the CALIFA data set used in this work. The grids of models (meta-grid and ad-hoc models for each region) are described in Sec.~\ref{sec:grid}. The results are presented and discussed in Sec.~\ref{sec:resdisc}, while the conclusions are drawn in Sec.~\ref{sec:concl}.

%__________________________________________________________________
\section{The CALIFA data}
\label{sec:califa}

The galaxies were selected from the CALIFA observed sample. Since CALIFA is an ongoing survey whose observations are scheduled on a monthly basis (i.e., dark nights), the list of objects increases regularly. The current results are based on the 612 galaxies observed up to June 2014, comprising both galaxies from the CALIFA mother sample and the so-called extended sample (details in  S\'anchez et al. in prep.). Their main characteristics have already been described in \cite{2015Sanchez_aap574}.

The details of the survey, sample, observational strategy, and
reduction are explained in \cite{sanchez12a}. All galaxies were
observed using PMAS \citep{roth05} in the PPAK configuration
\citep{kelz06}, covering a hexagonal field of view (FoV) of
74$\arcsec \times 64 \arcsec$, which is sufficient to map the full optical
extent of the galaxies up to two to three disk effective radii. This is
possible because of the diameter selection of the sample 
\citep{2014Walcher_aap569}. The observing strategy guarantees  complete coverage
of the FoV, with a final spatial resolution of FWHM$\sim$2.5$\arcsec$ \citep{rgb15},
corresponding to $\sim$1 kpc at the average redshift of the
survey. 
The sampled wavelength range and spectroscopic resolution
(3745\AA-7500\AA, $\lambda/\Delta\lambda\sim 850$, for the low-resolution
setup) are more than sufficient to explore the most prominent ionized
gas emission lines from \oii$\lambda$3727 
to \sii$\lambda$6731 at the redshift of our targets, on
one hand, and to deblend and subtract the underlying stellar
population, on the other \citep[e.g.,][]{sanchez12a,kehrig12,cid-fernandes13,cid-fernandes14}. The dataset was
reduced using version 1.5 of the CALIFA pipeline, whose modifications
with respect to the ones presented in \cite{sanchez12a} and \cite{husemann13}  are described in detail in \cite{rgb15}. In summary, the data fulfill the
predicted quality-control requirements with a spectrophotometric
accuracy that is better than 5\%\ everywhere within the wavelength range,
both absolute and relative with a depth that allows us to detect
emission lines in individual \HII\ regions as faint as
$\sim$10$^{-17}$\flux, and with a signal-to-noise ratio of
S/N$\sim$3-5. For the emission lines considered in the current study,
the S/N is well above this limit, and the measurement errors are
negligible in most of the cases. In any case, they have been
propagated and included in the final error budget.

The final product of the data reduction is a regular-grid datacube,
with $x$ and $y$ coordinates that indicate the right ascension and
declination of the target, and $z$ is a common step in
wavelength. The CALIFA pipeline also provides  the propagated error
cube, a proper mask cube of bad pixels, and a prescription of how to
handle the errors when performing spatial binning (due to covariance
between adjacent pixels after image reconstruction). These datacubes,
together with the ancillary data described in \citet{2014Walcher_aap569}, are the basic starting points of our analysis.

\subsection{\HII\ regions: detection and extraction}\label{HIIreg}

The segregation of \ion{H}{ii} regions and the extraction of the
corresponding spectra is performed using a semi-automatic procedure
named {\sc HIIexplorer}\footnote{\url{http://www.caha.es/sanchez/HII_explorer/}}. The
procedure is based on some basic assumptions: (a) \HII\ regions
are peaky and isolated structures with a strong ionized gas emission, which is
significantly above the stellar continuum emission and the average ionized gas
emission across the galaxy. This is particularly true for 
\Ha\ because (b) \HII\ regions have a typical physical
size of about a hundred or a few hundred parsecs
\citep[e.g.,][]{rosa97,lopez2011,oey03}, which corresponds to a typical
projected size of a few arcsec at the  distance  of the galaxies.

These assumptions are based on the fact that most of the \Ha\
luminosity observed in spiral and irregular galaxies is a direct
tracer of the ionization of the interstellar medium (ISM) by the
ultraviolet (UV) radiation produced by young high-mass OB
stars. Since only high-mass, short-lived stars contribute
significantly to the integrated ionizing flux, this luminosity is a
direct tracer of the current star-formation rate (SFR), independent of the
previous star-formation history. Therefore, clumpy structures detected in
the \Ha\ intensity maps are most probably
associated with classical \HII\ regions (i.e., those regions for which the oxygen
abundances have been calibrated).

The details of {\sc HIIexplorer} are given in \cite{sanchez12b} and
\cite{rosales12}. In summary we create a narrow-band image centered on
the wavelength of \Ha\ at the redshift of the object. Then we run
{\sc HIIexplorer} to detect and extract the spectra of each individual
\HII\ region, adopting the parameters presented in
\cite{sanchez14}. The algorithm starts looking for the brightest
  pixel in the map. Then, the code aggregates the adjacent pixels until all
  pixels with flux greater than 10\% of the peak flux of the region and
  within 500 pc or 3.5 spaxels from the center have been
  accumulated. The distance limit takes the typical size
  of \HII\ regions of a few hundreds of parsecs into account
  \citep[e.g.,][]{rosa97,lopez2011}. Then, the selected region is
  masked and the code keeps iterating until no peak with flux exceeding the median \Ha\ emission flux of the galaxy is left.
  \cite{mast14} studied the loss of resolution in IFS using nearby
  galaxies observed by PINGS \citep[PPAK ISF Nearby Galaxies Survey, see][]{rosales-ortega10}. Some of these
  galaxies were simulated at higher redshifts to match the
  characteristics and resolution of the galaxies observed by the CALIFA
  survey. Regarding the \HII\ region selection, the authors conclude
  that at $z\sim$0.02, the \HII\ clumps can contain on average from 1 to
  6 of the \HII\ regions obtained from the original data at
  $z\sim0.001$. Another caveat is that this procedure tends to select
  regions with similar sizes, although real \HII\ regions actually have
  different sizes. However, the actual adopted size is close to
  the FWHM (2.5$\arcsec$) of the CALIFA data for the version of the data
  reduction we used \citep{rgb15}.

Then, for each individual extracted
spectrum we modeled  the stellar continuum using
{\tt FIT3D}\footnote{\url{http://www.astroscu.unam.mx/~sfsanchez/FIT3D/}}
, a fitting package described in \cite{sanchez06b}, \citet{2011Sanchez_mnra410} and \cite{sanchez15c}. This fitting tool performs multiple linear regressions to derive the optimal combination of a single-stellar population (SSP) library over a set of Monte-Carlo realizations of the input spectrum, providing with best fitted set of weights for each population and the corresponding errors. Prior to this analysis the procedure derive the best kinematics and dust attenuation for each fitted spectrum. 
In this particular study we use the {\tt gsd156} template library,
described in detail by \citet{cid-fernandes13}. It
comprises 156 templates that cover 39 stellar ages (1 Myr to 13 Gyr),
and 4 metallicities ($Z/Z_{\odot}=$ 0.2, 0.4, 1, and 1.5). These
templates were extracted from a combination of the synthetic stellar
spectra from the GRANADA library \citep{martins05} and the SSP library
provided by the MILES project \citep{miles, vazdekis10, falc11}. This
library has been extensively used within the CALIFA collaboration in
different studies \citep[e.g.][]{eperez13, cid-fernandes13,
  rosa14}. The only difference with respect to these studies is that
the spectral resolution of the library was not fixed to the spectral
resolution of the CALIFA V500 setup data (FWHM$\sim$6 \AA), to allow
its use for datasets with different resolution.
As shown in \citet{sanchez15c} the results are not strongly affected by
the selection of a different stellar template.
The reliability of the derived parameters for the stellar population using {\sc FIT3D} was extensively tested against simulations and perturbed data. In particular it was found that it is required a S/N above 50 to break the well-known degeneracies and provide reliable weights for the stellar population when they contribute at least to $\sim$5\% to the total flux in the visible range \citep[e.g.][Fig. 9 and 15]{sanchez15c}. Indeed in previous articles \citep{sanchez2014a} we already explored the correspondence between the estimated fraction of young stars ($f_y$) and the equivalent width of H$\alpha$ (\EWa), parameters that show a clear correlation when the $f_y>$20\% and \EWa$>$6\AA. This is indeed a good test that supports the reliability of the derived fraction of young stars.

After subtracting the underlying stellar population, the flux intensity
of the strong emission lines was extracted for each gas-pure spectrum
by fitting a single Gaussian model to each line,
resulting in a catalog of the emission-line properties \citep{sanchez12b}. %Finally, the \HII\ regions were selected from this final catalog based on the
%properties of the underlying stellar continuum. 
% SFS NOTE: This is if we just want the regions dominated by young
% stellar populations. We have to remove that sentence in this context.

The final catalog comprises the strongest emission line and emission
line ratios from \oii$\lambda$3727 to
\sii$\lambda$6731 for 18178 \HII\ regions from 612 galaxies, together with their
equivalent widths and the luminosity-weighted ages and metallicities
of the underlying stellar population. So far, this is the largest
catalog of \HII\ regions and aggregations with spectroscopic
information. It is also one of the few catalogs derived for a
statistically well-defined sample of galaxies representative of the
entire population of galaxies in the local Universe \citep{2014Walcher_aap569}.

In this work, we will only use the HI, \nii, \oii, and \oiii\footnote{In the following, we will use \nii, \oi, \oii, \oiii, \sii, and \siii\ for the \nii$\lambda$6584\AA, \oi$\lambda$6300\AA, \oii$\lambda$3726+29\AA, \oiii$\lambda$5007\AA, \sii$\lambda$6716+31\AA, and \siii$\lambda$6312\AA\ lines respectively.} emision lines, which are present in almost all the sources. Auroral lines are seen only in some tens of them \citep[see the 16 \oiii$\lambda$4363 lines used by][]{2013Marino_aap559}. 

%\vspace 

\section{The grid of models}
\label{sec:grid}
\subsection{The meta-grid}\label{sec:meta_grid}

\begin{figure}
\centering 
\includegraphics[width=\hsize, trim = 20 0 20 0, clip = yes]{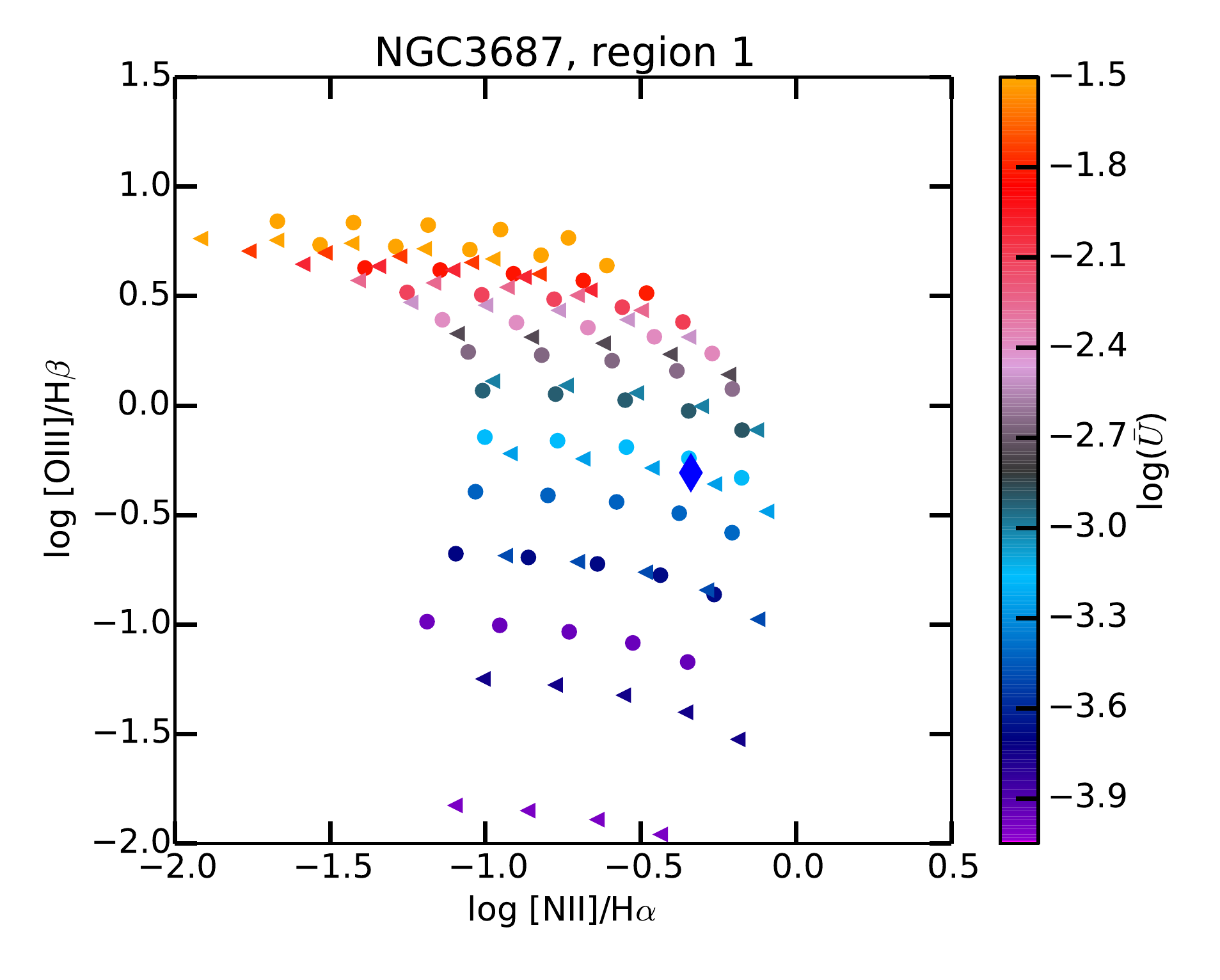}
\caption{Illustration of the modeled BPT diagram \oiii/\Hb\ vs. \nii/\Ha\ used to interpolate the values of \logU\ and N/O for the region 1 of NGC3687. The blue diamond corresponds to the observed values in this \oiii/\Hb\ vs. \nii/\Ha\ diagram. The circles and triangles correspond to the values of the models obtained with the morphological factor $fr$ set to 0.03 (filled sphere) and to 3.0 (thin shell) respectively. The colors correspond to the values of \logU\, while the different values of N/O lead to models from left to right for increasing N/O.}
\label{fig:BPT_1}%
\end{figure}

For each \HII\ region of each galaxy, we run a grid of 220 photoionization models using the {\sc Cloudy} code \citep[c13.03]{2013Ferland_rmxa49} driven by the pyCloudy package\footnote{\url{https://sites.google.com/site/pycloudy/}}  
\citep{2013Morisset_, 2014Morisset_}. The grids are obtained by varying the mean ionization parameter \logU \footnote{The ionization parameter is defined as $U(r) = {Q({\rm H}^0)}/{4.\pi.r^2.N_H.c}$, where $Q({\rm H}^0)$ is the number of ionizing photons emitted by the source per unit of time, $r$ is the distance between the source and the nebula, $N_H$ is the hydrogen density and $c$ is the speed of light. We use the mean value of $U$ on the volume of the nebula weighted by the electron density, and name \logU\ its logarithmic value.} which takes 11 values from -4 to -1.5, the abundance ratio log(N/O) (5 values from -1.5 to 1.5 around the solar value), the morphology parameter $fr$ (0.03 and 3.0, see below) and the nebular metallicity ("Neb" and "Stel", see below).
Following \citet{2015Stasinska_aap576}, the desired value of \logU\ is obtained by setting the H$^0$-ionizing photons emission rate $Q(\rm H^0)$ to:

$$Q({\rm H}^0) = \frac{4 \times \pi \times c^3 \times \bar U^3}
					{3 \times N_H \times ff^2 \times \alpha_B^2 \times w^3}$$

where $c$ is the speed of light, $N_H$ is the hydrogen density (set to 10 H/cm$^3$ for all the models), $ff$ is the filling factor, $\alpha_B$ is the effective case B recombination coefficient, and $w = (1 + fr^3)^{1/3} - fr$, with the morphology factor $fr = R_{in}/R_{Str}$ being the ratio between $R_{in}$, the inner radius of the nebula, and $R_{Str}$, the Strömgren radius of the nebula if it where a full filled sphere. A morphology factor $fr >> 1$ ($w\sim 0$ ) corresponds to a thin shell (e. g. a plan parallel model), while $fr\sim 0$ ($w\sim 1$ ) corresponds to a filled sphere. The Strömgren radius of a filled sphere is:

$$R_{Str} = \left[ \frac{3 \times Q({\rm H}^0)}{4 \times \pi \times N_H^2 \times \alpha_B \times ff}\right]^{1/3}$$

The ionizing SED is obtained by summing up individual models from POPSTAR code \citep{2009Molla_mnra398}. Each model corresponds to an individual burst of age and metallicity from the {\tt gsd156} template and has a weight given by the multi-SSP analysis described before. We use POPSTAR models obtained following the IMF from \citet{Salpeter:1955p3438}. We checked that using an IMF from \citet{Chabrier:2003p3777} does not significantly change our results.

The oxygen abundance of the ionized gas is determined in two ways:
\begin{itemize}
\item "Neb": from the nebular \oiii/\nii\ line ratios applying the O(O3N2) relation determined by \citet[][hereafter M13]{2013Marino_aap559}, namely 12 + log (O/H) = 8.533 - 0.214 $\times$ O3N2, where O3N2 is $\log$((\oiii/\Hb)/(\nii/\Ha)).

\item "Stel" : The luminosity-weighted log metallicity of the underlying young stellar population, derived by co-adding the metallicities of the corresponding SSPs within the library multiplied by its contributed fraction of light in the $V$-band, but only for those SSPs with ages younger than 2~Gyrs, following \citet{rosa14}. This set of models will not be used in the following main analysis, as we know that this determination of the nebular metallicity is less reliable, due to the low S/N of the underlying continuum for a fraction of the\HII\ regions \citep{sanchez15c}. Although the continuum has a good S/N ($\sim$30-50) to perform a SSP decomposition, to derive the metallicity of the young stars it is needed to have a similar S/N for only the young component, that may contribute between a 100\% and a 20\% of the total flux. This can not be granted in general. Therefore, the estimated metallicities would have large errors. The results obtain with these models are discussed in \ref{sec:stelabund} and are showed in the Online only Appendix \ref{append:stelabund}. This method have been calibrated to be used for \HII\ regions and may not apply for regions ionized by old stars.

\end{itemize}

Notice that given the range of masses of the galaxies considered here (10$^{9.5}$ to 10$^{12}$ solar masses), we expect the O/H abundance to range from 10$^{-4}$ to 10$^{-3}$, based on the Mass-Metallicity relation \citep[e.g][]{sanchez13}.

The element abundances are following O/H (except N/H which is a free parameter). The abundances relative to O are taken from \citet{2009Asplund_araa47}.

Dust is included in the model, in the form of the "ism" type defined by Cloudy, with a dust to gas ratio following a broken power law, as in the case X$_{\rm CO, Z}$ defined and recommended by \citet{2014Remy-Ruyer_aap563}. Following \citet{2011Draine_apj732}, we apply an additional factor of 2/3 to the final dust to gas ratio used.

The models of this meta-grid (grid of grids) are run in a quick mode (no iteration, no level-2 lines; see Cloudy manual). The ad-hoc models (see next section) do not have these limitations.

Not all the CALIFA regions have been used, we apply a filter to select only the ones where the values of the CALIFA field ratio med\_flux/StdDev are over 10 for the lines of interest (this correspond to the average S/N from blue to red through the full spectral range) and the value of the CALIFA field MIN\_CHISQ is over 0.7 (the reduced chi$^2$ parameter). Both cuts ensure that the fitting provides reliable results. A total of 397 galaxies have been used, summing up 9181 regions corresponding to 2,019,820 individual photoionization models. 

The models have been stored in the unpublished working database associated with 3MdB \citep[Mexican Million Models database, see][]{2015Morisset_rmxa51}.

\subsection{The ad-hoc models}

We use the meta-grid to find the model that reproduces the best the observed line ratios \oiii/\Hb\ and \nii/\Ha\ for each region. Figure~\ref{fig:BPT_1} illustrates as an example the results for region 1 of NGC3687. The figure shows the classical BPT diagram, \oiii/\Hb\ vs. \nii/\Ha\ \citep{1981Baldwin_pasp93}. The blue diamond corresponds to the observed line ratios whereas the colored circles (triangles) are the results of the grid of models obtained with $fr$ being 0.03 (3.0), using the SED corresponding to this region and the nebular metallicity derived from the O3N2 diagnostic (see Sec.~\ref{sec:meta_grid}).

We have performed a 2D-interpolation in the BPT diagram to determine the \logU\ and the N/O ratio for each of the models. The metallicity method ("Stel" or "Neb", see sec.~\ref{sec:meta_grid}) and the geometry ($fr$) take both two values and thus, we obtained 4 ad-hoc models for each region. As no extrapolation is done in the BPT diagrams, for some regions less than 4 ad-hoc models are performed. The final number of ad-hoc models is 20,793 (from 272 galaxies), from which 10,196 are "Neb" models (obtained using the O3N2 method to determine O/H).

All the 20,793 ad-hoc models are stored in the 3MdB database \citep{2015Morisset_rmxa51} and accessible for any user under the "CALIFA\_ah" reference (the "ref" field). The "com5" field is used to store the value of the fit for the \oii/\Hb/ line ratio, see Sec.\ref{sec:O2filter}. The "com8" field is used to store the result of the BPT-Population filter, see Sec.~\ref{sec:BPT-pop}.
Once the final release of the CALIFA data will be publicly available, we plan to rerun all the procedure to obtain more data points and to store the corresponding new models in 3MdB under the "CALIFA\_ah2" reference.

\subsection{The \oii/H$\beta$ filter}
\label{sec:O2filter}
The ad-hoc models have been designed to fit the \nii/\Ha\ and \oiii/\Hb\ line ratios, but we can also impose that they fit the observed \oii/\Hb\ line ratio. We filter the models for which the value of $\log$(\oii/\Hb) equals the observed value within 0.1 dex. That reduces the number of models by a factor of $\sim6$ but provides us with a more realistic reference data set. To apply this filter we have corrected the \oii\ line intensities from the reddening, using \Ha/\Hb\ = 2.85 and the \citep{1999Fitzpatrick_pasp111} extinction law. The number of ad-hoc models that also fit this \oii/\Hb\ filter is 3195, from which 1574 are "Neb" models. In the 3MdB database, one can select the models that fit the \oii/H$\beta$ line ratio, using the "com5" entry, that contains log(\oii/H$\beta$)$_{obs}$ - log(\oii/H$\beta$)$_{mod}$.

This filter certainly adds some bias in our sample, as only the 1/6 of the models remains.

Note that other emission line ratios could also be used to filter models, but they all involve abundance ratios that are not free parameters in our modeling process (e.g. using \siii/\Hb\ depends on the S/H abundance ratio). In other words, not fitting the observed \siii/\Hb\ ratio for a given observation may only indicates that the S/H abundance is not correct; but S/H is determined by fixing S/O, we have no way to act on \siii/\Hb\ ratio. The fact that S/H may be incorrect has virtually no consequences on the results presented in the following sections, but this would artificially exclude "good" models if we use the \siii/\Hb\ ratio as a filter.

\subsection{Characterizing the ionizing population}\label{sec:Pop}

\begin{figure}
\centering 
\includegraphics[width=\hsize]{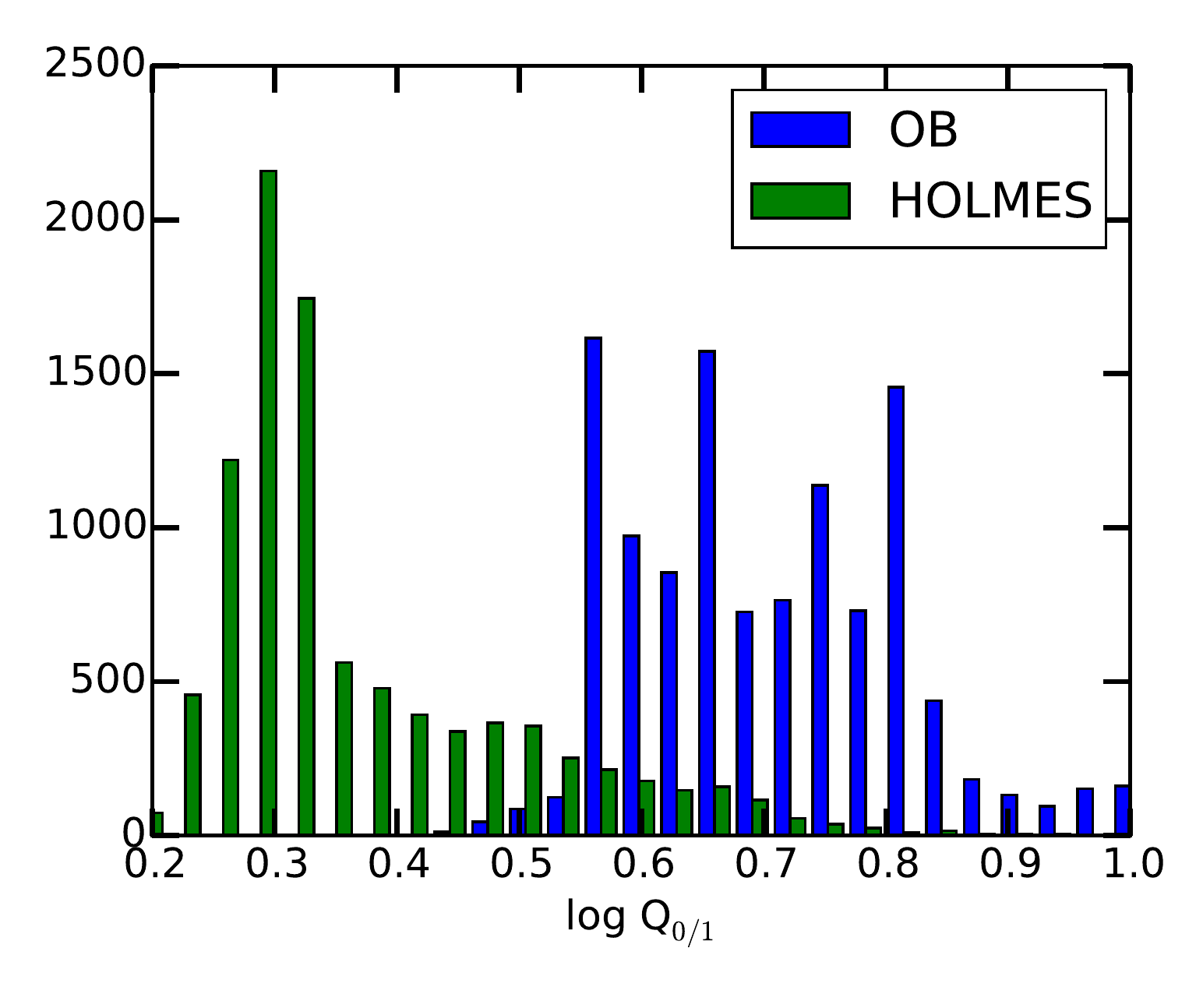}
\caption{Histogram of \QHHe\ = Q(H$^0$)/Q(He$^0$) for the OB and HOLMES dominated spectra.}
\label{fig:pop_hist}%
\end{figure}

From the study by \citet{2015Morisset_rmxa51}, we can define limits in an age-metallicity plane for the ionizing stellar populations. Using their Fig.~5, we can determine that, for log(O/H) < -3.5, OB stars correspond to log(age/yr) < 6.8 and HOLMES correspond to log(age/yr) > 8.25. For log(O/H) > -3.5, OB stars correspond to log(age/yr) < 6.7 and HOLMES correspond to log(age/yr) > 7.9. Using these limits and the decomposition of the spectra on the {\tt gsd156} library base, we can define what is the proportion of the ionizing photons coming from OB stars in the total number of ionizing photons Q0. This proportion will be named \fOB\ in the rest of the paper. There is a strong correlation between the type of dominant ionizing stellar population depicted by this \fOB\ and the value of \QHHe\ = Q(H$^0$)/Q(He$^0$), the ratio of the number of photons ionizing H$^0$ and He$^0$ (a kind of softness parameter). This is illustrated by Fig.~\ref{fig:pop_hist}, where the histograms of \QHHe\ obtained for OB stars and HOLMES are compared. There is a clear separation at a value of $\sim 0.55$, HOLMES being associated with the lowest values.
In the following we will mainly use the \QHHe\ ratio to trace the underlying population, keeping in mind that the purple/bluish dots point to HOLMES and the gray/reddish/yellow dots to OB stars for all figures throughout this article. 

\subsection{The H$\alpha$ equivalent width}
\label{sec:HaEW}

\begin{figure}
\centering 
\includegraphics[width=\hsize]{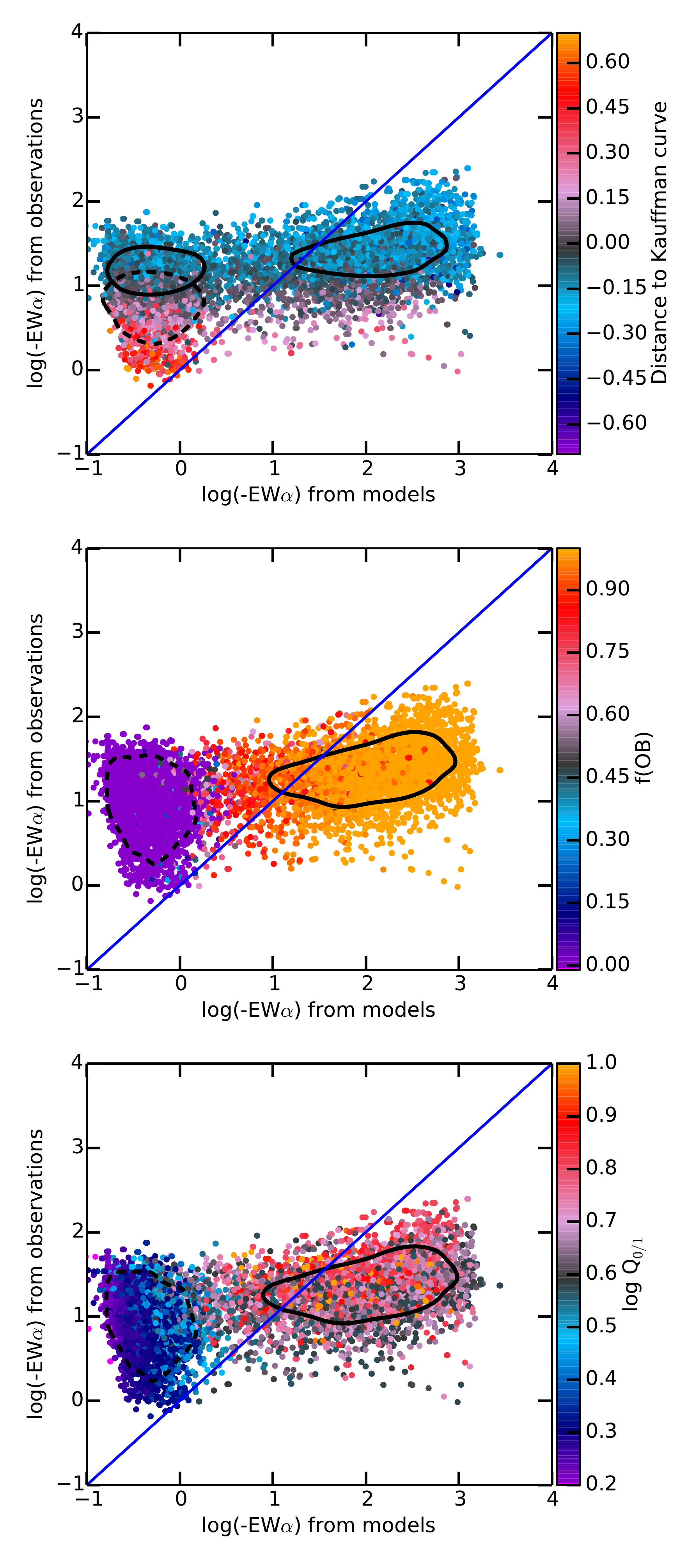}
\caption{Comparison between the \EWa\ from the models and from the observations. Colors are representing K$_{dist}$ , the distance to the \citet{2003Kauffmann_mnra346} curve (upper panel), the OB stars proportion \fOB\ (middle panel), and the ratio \QHHe\ = Q(He$^0$)/Q(H$^0$) (lower panel). Solid blue line follows y=x. In the upper panel, the solid black curves enclose half of the models which are below the Kauffman curve (negative K$_{dist}$), while the dashed black curve encloses half of the models above the same curve (positive K$_{dist}$). In the middle panel, the solid black contour encloses half of the models ionized by OB stars (\fOB\ > 0.5), while the dashed black contour encloses half of the models ionized by HOLMES (\fOB\ < 0.5). In the lower panel,  the solid black contour encloses half of the models ionized by OB stars (\QHHe\ > 0.55), while the dashed black contour encloses half of the models ionized by HOLMES (\QHHe\ < 0.55).}
\label{fig:EW_comp}%
\end{figure}

We can compare the observed \Ha\ equivalent widths (\EWa) with the prediction from the models. We first corrected the observed value from the extinction, using \Ha/\Hb\ = 2.85 and the \citet{1999Fitzpatrick_pasp111} extinction law to correct the \Ha\ flux and using the A$_V$ from the stellar observations to correct the stellar continuum. It is well known that the stellar continuum is affected by less dust attenuation than the ionizing gas, in general, for star-forming galaxies \citep{2001Calzetti_pasp113}. The resulting differential correction has a median of 0.9 $\pm$ 0.1, changing virtually nothing in our results, but adding a few aberrant values due to bad observations of the \Ha/\Hb\ ratio. Finally we did not apply the correction.

\begin{figure}[!h]
\centering 
\includegraphics[width=\hsize]{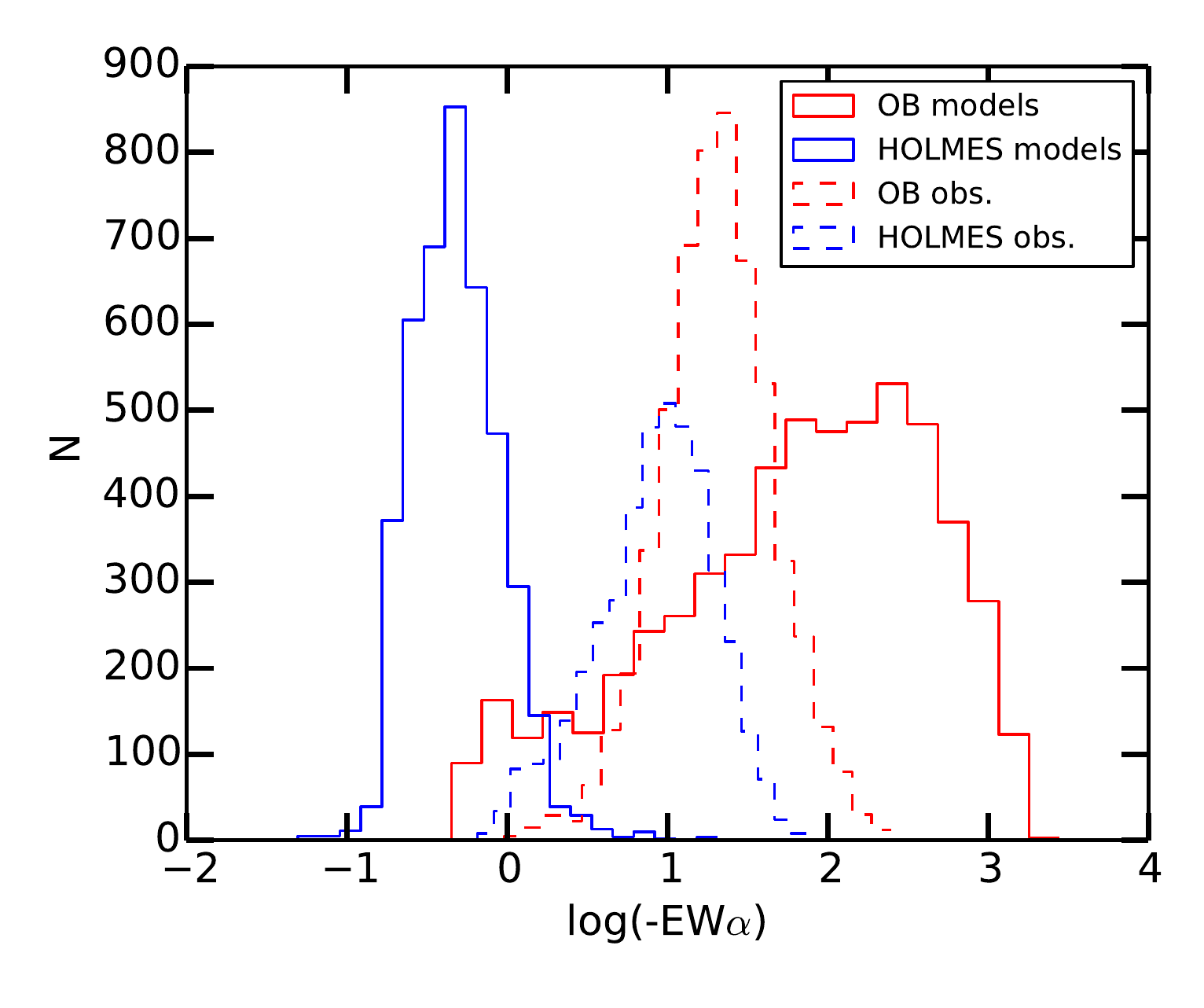}
\caption{Histogram of \EWa\ for the models (solid lines) and observations (dashed lines) for star forming regions (red, \QHHe\ > 0.55) and HOLMES ionized regions (blue, \QHHe\ < 0.55).}
\label{fig:EW_hist}%
\end{figure}

We plot in Figure~\ref{fig:EW_comp} the comparison between the \EWa\ from the observations and those from the computed models. Three color codes are used: the first one in the upper panel is associated to the distance to the Kauffman curve \Kdist\footnote{The \Kdist\ parameter is defined by the minimum value of the distance D between a given point and the \citet{2003Kauffmann_mnra346} curve, being negative for points below the curve and positive otherwise. The distance between 2 points in the BPT diagram is defined by D = $\sqrt{\Delta(\log(\oiii/\rm H\beta))^2 + \Delta(\log(\nii/\rm H\alpha))^2}$} 
(upper panel). 
In the middle panel \fOB\ is used as color code while in the bottom panel, the color code is following \QHHe.
The solid line in the plot represents where both values of \EWa\ are equal.

We plot in Figure~\ref{fig:EW_hist} the histograms of \EWa\ for the observation and the models, for the regions ionized by HOLMES (in blue) and by OB stars (in red). These histograms actually correspond to the bottom panel of Figure~\ref{fig:EW_comp}.

To our knowledge, this is the first time the two values of \EWa\ (observed and modeled) are compared for such a set of objects where a detailed determination of the underlying stellar population is available.

We can easily see that there is a clear trend in the variations of the colors in the middle and lower panels, indicating that the ratio \QHHe\ is a good indicator of the type of stars dominating the ionizing flux (see also Fig.~\ref{fig:pop_hist} in Sec.~\ref{sec:Pop}). The color code used for \QHHe\ reflect the fact that OB stars (red/orange) are the coolest of our sample, while HOLMES (the hottest) are in violet.

In the upper panel of Fig.~\ref{fig:EW_comp} we explore how the differences between the observed and modeled \EWa\ are related to the \Kdist\ parameter. We can see that most of the regions below the Kauffman curve (\Kdist < 0, blue regions), that correspond to classical HII regions, have observed values of \EWa\ > 10, while most of the regions above the same curve (HOLMES, low-ionization nuclear emission-line region, i.e. LINERS etc) have observed values of \EWa\ < 10. 

The distribution of the source of the ionizing photons described by \fOB\ and \QHHe\ are clearly bimodal: there is no example of regions ionized by OB stars in a proportion around 50\%. This means that we found a clear separation in the models between classical HII regions and regions ionized by old stars. Fig.~\ref{fig:EW_comp} also shows the contours enclosing the two different populations highlighted in each panels: the models falling above and under the Kauffman curve (positive and negative values of K$_{dist}$ resp., top panel) and the regions ionized by OB stars and by HOLMES (value of \fOB\ and \QHHe, middle and lower panels respectively). This indicates if these two populations are distinguishable using the values of \EWa .

The \EWa\ is related to the ratio between the number of ionizing photons actually processed by the gas and the number of (ionizing) stars. Both the gas and the stars are supposed to be included in the observed beam. We can see from the Fig.~\ref{fig:EW_comp} that there is an obvious trend between the observed and modeled \EWa. The regions lying on the right side of the $y=x$ line, thus having an observed \EWa\ lower than the modeled value, correspond to regions where less photons are ionizing the gas than what is expected in a closed geometry model, leading to what is commonly called "leaking". \citet{2001Stasinska_aap370} also proposed the presence of old populations to explain this discrepancy, but we include these old stars in our modeling process.
This leaking can be due to matter-bounded regions (in which there is not enough material to be ionized in some directions from the ionizing source point of view) or because of a covering factor less than one (in some directions there is no gas at all), or even a combination of both effects. In both cases, some ionizing photons escape the region. In the case of a covering factor less than one, the photoionization models correctly predict the line ratios (only the absolute fluxes are overestimated, all in the same proportions) contrary to the case of matter-bounded regions. The presence of strong \oi\ and \oii\ emission lines favors the idea of these \HII\ regions having a covering factor less than one, then validating the use of the photoionization models.

On the other side of the $y=x$ line (the left one), the \EWa\ from the observations is higher than the one from the models. This corresponds to regions where the amount of ionized gas is too much compared to what could be ionized by the observed stellar population. It means that we are missing some of the ionizing sources for these regions. It can actually be the effect of photons coming from sources out of the beam, perhaps escaping from the same regions previously described (located on the right side of the line). It would be very interesting to estimate the number of leaking photons in each galaxy and then see if they are enough to explain the discrepancies in the \EWa\ determined from observations and models, but this is out of the scope of this paper. 

The difference between the observed and the modeled \EWa\ is clearly related to the proportion of OB stars (top and middle panels of Fig.~\ref{fig:EW_comp}) or the hardness of the ionizing radiation (bottom panel). Most of the star forming regions are located on the right side of the $y=x$ line and correspond to photon leaking. On the other hand, the regions mainly photoionized by HOLMES are only on the left side and correspond to regions ionized by additional sources. The same conclusion can be reached analyzing Fig.~\ref{fig:EW_hist}, where observed star forming regions (red dashed line) have a median log(\EWa) of $\sim 1.3$, while the corresponding models have a median log(\EWa) of $\sim 1.9$, leading to a factor of leaking $\sim 4$ (80\% of the photons escape). This median leaking is the same if we use \EWb, as all the equivalent widths are shifted by $\sim -0.5$~dex. Notice that the distribution of the \EWa\ values for the models of \HII\ regions is broader than the distribution of the observed values, reflecting the variety of morphologies leading to leaking factors from 1.0 (no leaking) to some tens \citep[See also][]{2013Papaderos_aap555}.
When considering regions ionized by HOLMES (blue lines in Fig.~\ref{fig:EW_hist}), we see that the discrepancy between the observed and modeled values of  \EWa\ goes in the other direction (regions in the left part of the Figs.~\ref{fig:EW_comp}), with log(\EWa) median values of $\sim1.0$ and $\sim-0.3$ for observation and models respectively.

Previous studies have analyzed the nature of the LINER-like emission in galaxies \citep{2013Papaderos_aap555, sign13, sarzi10}. All of them concluded that the nature of this ionization is most probably due to post-AGBs stars, i.e., HOLMES in our nomenclature. In particular \citet{2013Papaderos_aap555} and \citet{gomes16} presented a comparison between the observed \Ha\ fluxes and the predicted ones based on photoionization models which ionizing source was selected from the analysis of the underlying stellar population for a sample of early-type galaxies. Thus, their analysis is somehow similar to the one presented here. They found that there are two kinds of galaxies on the basis of this comparison: type-i, for which the observed and predicted fluxes match very well, in general, and type-ii, for which they describe a deficit of observed flux, compatible with a Lyman-continuum leaking, in agreement with the results presented here. We need to recall that by selection procedure adopted we have excluded a substantial fraction of the diffuse regions, that are those ones dominating the type-i ETGs.

We must notice that state-of-the-art population spectral synthesis models are still plagued by significant degeneracies (e.g., the notorious age-metallicity degeneracy) and uncertainties in the best-fitting star formation history. A tiny variation/uncertainty in the mass fraction of young (<15-20 Myr) ionizing stars (simple stellar population-SSP models in this study) results in a very significant change in the expected value of Q$_0$, consequently the Balmer recombination line luminosities. On the other hand, the discrepancy between the observed and modeled \EWa\ in case of HOLMES can just come from an underestimation of the ionizing flux from post-AGBs. These kind of short-lived, highly variable period of the evolution of stars is not very well understood, and its inclusion in SSP models is quite recent, being still a topic that present large uncertainties. Even in the case that we derive correctly the fraction of SSPs comprising post-AGBs it may be still the case that the predicted ionizing photon distribution is not totally correct. For all those reasons we prefer not to include those ionized regions in further analysis (see next section), concentrating ourselves in the much better understood regions ionized by young stars. In further studies we will try to improve our analysis: (1) excluding or subtracting the possible contribution of a central AGN, if feasible, and (2) updating as much as possible the SSPs and the ionization models adopted for post-AGB stars. 

\subsection{The BPT-population filter}
\label{sec:BPT-pop}

The regions that correspond to star forming regions in the BPT diagram (negative K$_{dist}$ values) and that are on the left side of the y=x line in the Fig.~\ref{sec:HaEW} are not leading to trustable models, as the ionizing source should be of OB-stars type, and what is obtained from the SSP decomposition is of type HOLMES. We apply another filter to the ad-hoc models to remove objects that have negative K$_{dist}$ values and that are ionized by old populations. Applying this BPT-population filter to the ad-hoc models already filtered by \oii/\Hb\ leads to a set of 2558 models, from which 76\% are star forming regions, and 24\% are regions ionized by HOLMES and which are over the Kaufmann curve. In the 3MdB database, we set to 1 the value for the "com8" field for the models that fit this filter (and 0 otherwise).

The final 2558 models used in the next section fit the \nii, \oii, and \oiii\ lines with the following mean value and standard deviation: \nii/\hb\ Model/Obs = 1.05 $\pm$ 0.08, \oii/\hb\ Model/Obs = 0.98 $\pm$ 0.13, and \oiii/\hb\ Model/Obs = 1.04 $\pm$ 0.14.

In the following sections, the figures show the star forming regions and the ones ionized by HOLMES. But we select only star forming regions to compute {\bf fits} to our results, given that we are actually not sure about the pertinence of the HOLMES models (even after applying the filter described above): the missing photons may have a very different distribution, and the O/H abundance have been obtained using the O3N2 ratio, calibrated on \HII\ regions, not on HOLMES-ionized regions.

\section{Results and discussion}
\label{sec:resdisc}

In all the following sections, we will present results obtained using the ad-hoc "Neb" models selected after applying the [OII] filter described in Sec.~\ref{sec:O2filter} and the BPT-population filter described in Sec.~\ref{sec:BPT-pop}.

\subsection{A set of models nearly compatible with the direct method}

\begin{figure*}[!h]
\centering
\includegraphics[width=\hsize, trim = 30 10 100 10, clip = yes]{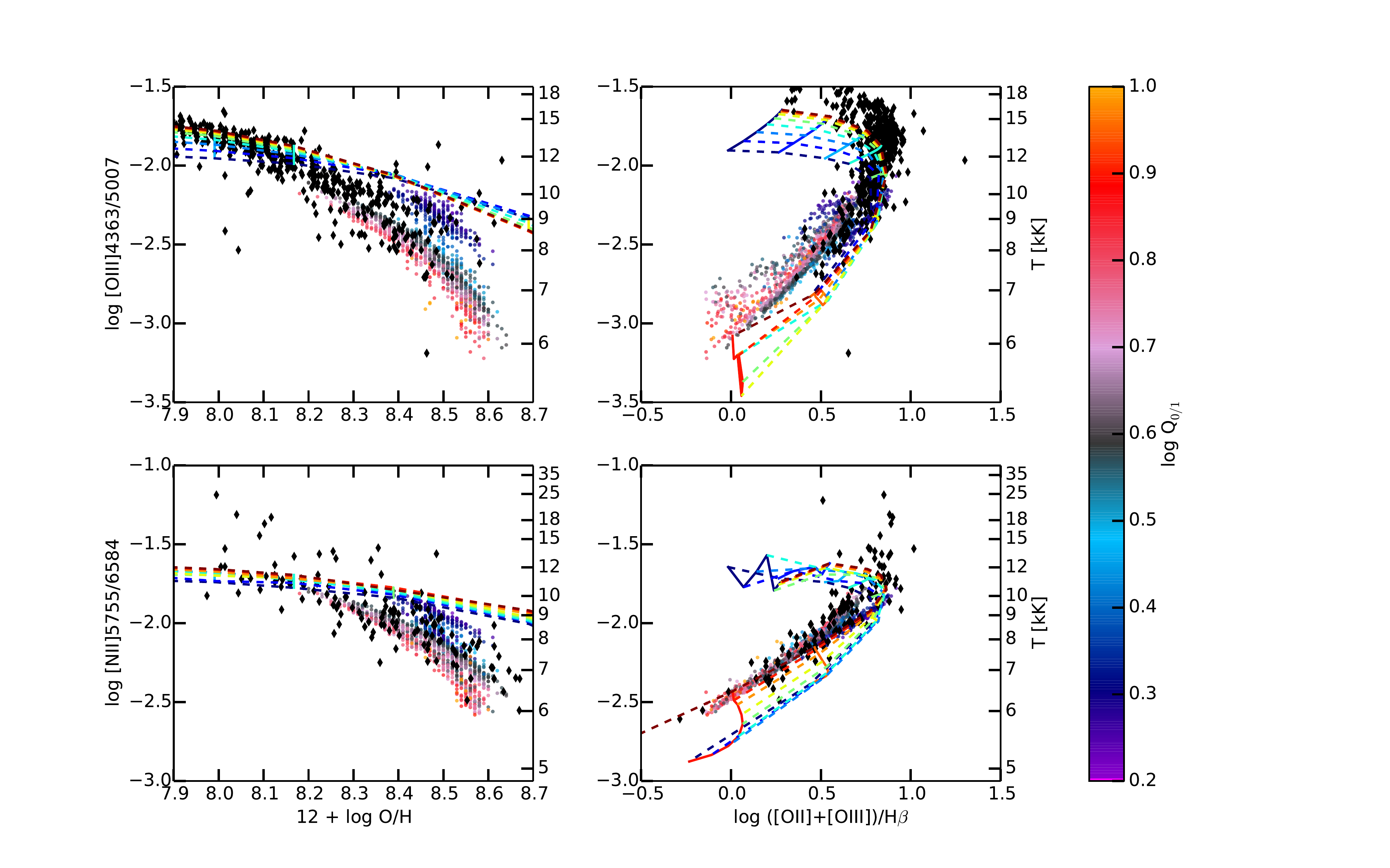}
\caption{Electron temperature diagnostic line ratios as a function of O/H (left panels) and (\oiii$\lambda$5007 + \oii$\lambda$3727)/H$\beta$ (right panels) for our models (colored circles). From top to bottom, the diagnostics are \oiii$\lambda$4363/$\lambda$5007 and \nii$\lambda$5755/$\lambda$6584. Black diamonds represent the \temp-based sample of \HII\ regions used by \citet{2013Marino_aap559}. The dashed-color lines correspond to the grid of models computed by \citep{2013Dopita_apjs208}. The color bar is following \QHHe, the softness of the ionizing radiation.}
\label{fig:comp_Te}%
\end{figure*}

In Fig.~\ref{fig:comp_Te}, we show the values of two \temp-diagnostic line ratios as a function of the oxygen abundance (left panels) and the (\oii$\lambda$3727 + \oiii$\lambda$5007)/\Hb\ line ratio (right panels) for the photoionization models (only the "Neb" models, colored circles), along with one of the largest compilation of \HII\ regions up-to-date with O/H derived using the direct-method (M13) and
for a set of observations used by \citet{2013Marino_aap559} (black diamonds). Top and bottom panel are showing the respective \oiii$\lambda$4363/5007 and \nii$\lambda$5755/6584 diagnostics. 

We find a good agreement between the models and the observations in all the panels, for metallicities corresponding to 12+log(O/H) > 8.2. Our star forming region models (orange/red/purple models) are falling over the coolest observed regions, pointing to a small underestimation of the electron temperature (but see below). The HOLMES\footnote{HOLMES stands for HOt Low Mass Evolved Stars, see e.g. \citet{2011Flores-Fajardo_mnra415}}-ionizing models (turquoise/blue) models are falling over the hottest regions.
This indicates that the set of models computed for this work, for which the O abundances have been calculated with the O3N2 method, is globally compatible with the determination of O/H using a direct method. We check that the "Stel" models are also falling on the same regions in these diagrams, leading to the conclusion that this behavior is not due to the way we define O/H. Other previous sets of photoionization models systematically show discrepancies between the values used as input for the O abundances and the values determined from the direct method \citep[e. g.][]{2012Lopez-Sanchez_mnra426}. This is illustrated with MAPPINGS models presented in the Fig.~\ref{fig:comp_Te} with a grid that shows the model results obtained from the tables electronically published by \citet{2013Dopita_apjs208}.
All these O/H values are systematically larger than the ones obtained with the direct method (black diamonds), while the results for our models cover the same region and describe the same trend. If we consider only star forming regions, our models are a little bit too cold, as they reproduce only half of the observed values of \oiii$\lambda$4363/5007.
As far as we know this is the first time that it is possible to reconcile the predictions by photoionization models with at least a significant amount of direct estimations of the abundance and the line ratios.

On the right panels we can see the differences between the models obtained for this work and the \citet{2013Dopita_apjs208} models obtained with MAPPINGS. While our models show a very good agreement with the observations, we can see in the upper and middle panels that some of the MAPPINGS models fall in a region where no observations are found (they turn around the data points cloud).

The main differences between the two sets of models seems to be the \temp\ obtained for a given value of O/H, depicted for example by a difference of $\simeq 0.5$~dex in \oiii$\lambda$4363/5007 at log(O/H)=-3.5, MAPPING models being hotter. This is probably the result of a lower heating or a higher cooling in our models. A lower heating can be explained by the use by \citet{2013Dopita_apjs208} of Starburst99 models \citep{1999Leitherer_apjs123} from 2005, which provides a harder radiation field than the newest models. In contrast, we use in our models the SED predicted by POPSTAR models \citep{2009Molla_mnra398}. 

\begin{figure}[!h]
\centering
\includegraphics[width=\hsize]{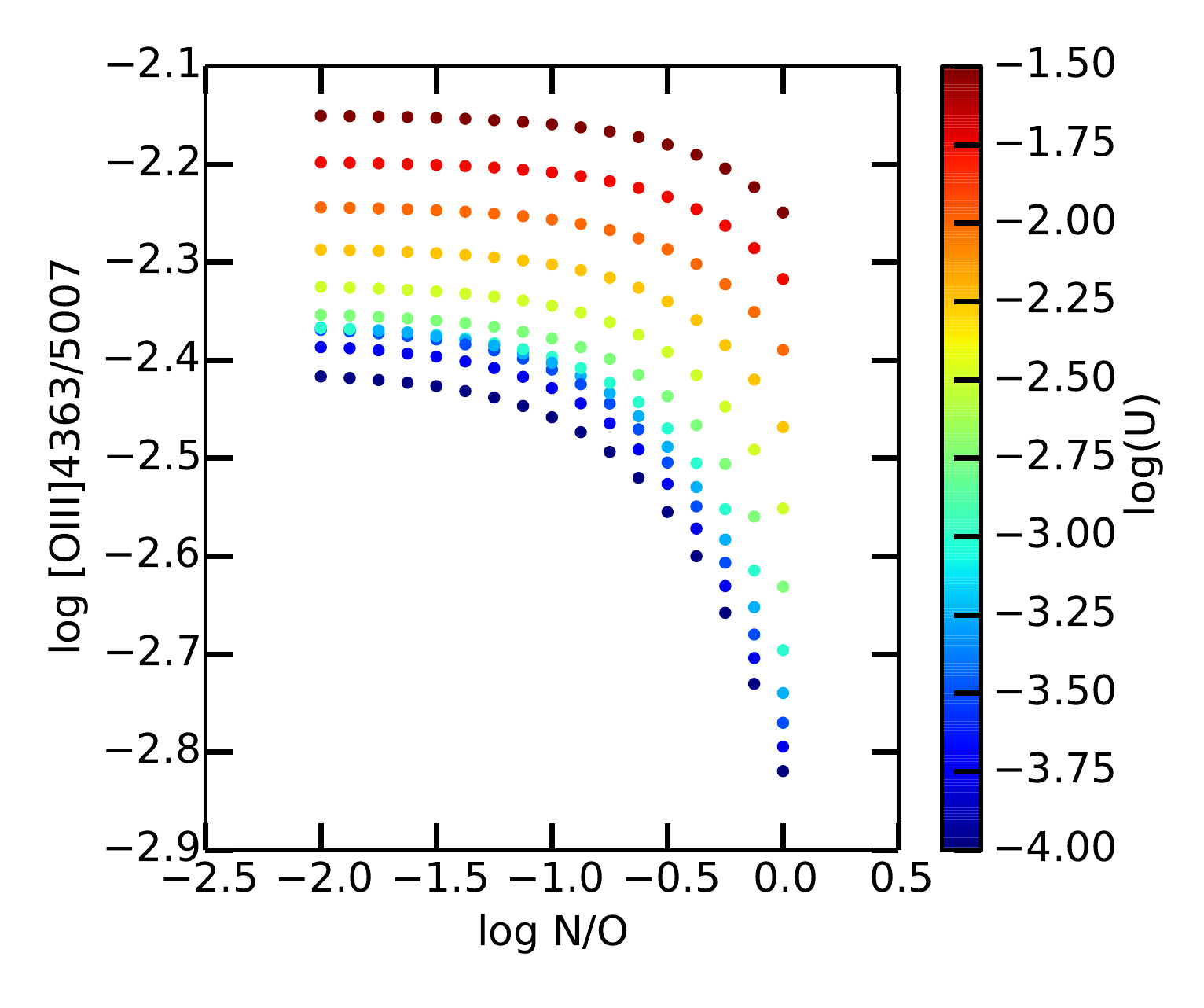}
\caption{Electron temperature diagnostic line ratios \oiii$\lambda$4363/5007 vs. N/O. The color bar is indicating \logU. The models are extracted from 3MdB and correspond to "HII\_CHIm" models of 1~Myr HII regions with log(O/H)=-3.5.}
\label{fig:comp_Te_NO_logU}
\end{figure}

A higher cooling can be due to our higher values of N/O for a given O/H (see Sec.~\ref{sec:NOoOH}), our lower values of \logU\ for a given O/H (see Sec.~\ref{sec:logUOH}) or the fact that we do not consider depletion of some elements. To test the effect of N/O and \logU\ on the electron temperature, we extract a subset of models from 3MdB \citep{2015Morisset_rmxa51}. We use the "HII\_CHIm" models \citep{2014Perez-Montero_mnra441} with log(O/H) = -3.5 and an age of the ionizing stellar cluster of 1~Myr, N/O and \logU\ let free. We show in Fig.~\ref{fig:comp_Te_NO_logU} both effects on the line ratio \oiii$\lambda$4363/5007 of changing N/O (on the x-axis) and \logU\ (the color code).
We can see that a difference in log(N/O) from -0.8 to -1.5 implies a very small difference on the T$_{\rm e}$-diagnostic line ratio ($\simeq 0.05$~dex). Changing \logU\ from -2.5 to -3.5 leads also to a very small effect on \oiii$\lambda$4363/5007 ($\simeq 0.05$~dex). 
To test the effect of the depletion of some elements on the electron temperature, we run two models with 1) the abundances from \citet{2009Asplund_araa47} as used in our models, and 2) a depletion of 1~dex for Si and Mg and of 1.5~dex for Fe. The effect on the line ratio \oiii$\lambda$4363/5007 is of increasing it by 0.09~dex, not enough to explain the observed difference of $\simeq 0.5$~dex at log(O/H) $\simeq$ -3.5, but almost enough to increase the ratio to the region where the observations are. 

We conclude that the differences observed in Fig.~\ref{fig:comp_Te} (especially in the upper-left panel) between our grid of models and the models from \citet{2013Dopita_apjs208} are neither due to the differences in N/O, or the differences in \logU\ nor the use or not of depletion. It may reside in the choice of the ionizing SED, if not coming from the code used. In summary, our photionization models are almost compatible with the electron temperature derived from the direct method at a given O/H, being the first ones in the literature to our knowledge. Our electron temperature is a little bit cool, perhaps pointing to a small lack of depletion of some elements or the presence of some light extra heating process or the presence of a process favoring the emission of high temperature lines (temperature fluctuations {\it a la} \citet{1967Peimbert_apj150}, $\kappa$ distribution {\it a la} \citet{2012Nicholls_apj752}). 
Indeed, the differences in the selected ionizing source could explain and solve the long standing incompatibility between the direct method and photoionization models, and we are presenting in this work a set of models approaching the observational reality.

\subsection{The BPT diagram}
\begin{figure*}[!h]
\centering
\includegraphics[width=\hsize]{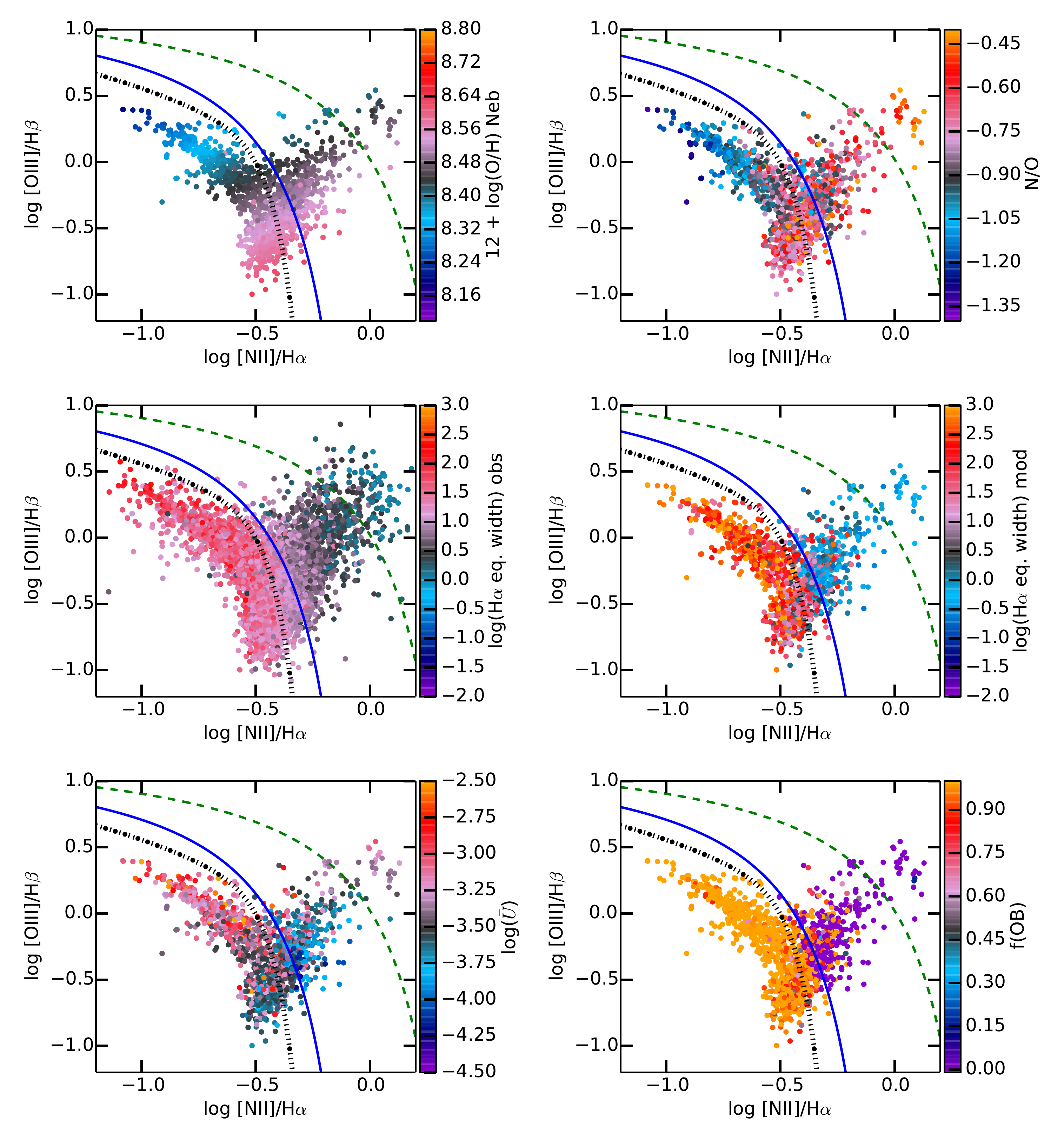}
\caption{Classical BPT diagrams of the model results. For each panel, the color code is changed according to the description on the right of the corresponding color bar. Upper panels show distributions of the models, with colors related to the chemical abundances: O/H on the left and N/O on the right. The middle panels show the distribution of the \Ha\ equivalent width, determined from the observation on the left and from the models on the right panels respectively. The lower panels show the distribution of the mean ionization parameter \logU\ on the left and the proportion of OB stars in the ionizing SED on the right panels respectively. The solid blue line is from \citet{2003Kauffmann_mnra346}, the green dashed line is from \citet{kewley01} and the dotted black line is from \citet{2006Stasinska_mnra371}.}
\label{fig:BPT_2}%
\end{figure*}

Figure~\ref{fig:BPT_2} shows our photoionization models in the classical BPT diagram \citep{1981Baldwin_pasp93}, $\log$(\oiii/\Hb) versus $\log$(\nii/\Ha). The curves derived by \citet{2003Kauffmann_mnra346}, \citet{2002Kewley_apjs142} and \citet{2006Stasinska_mnra371} have been included in the plots for reference. These curves are often used to distinguish between star-forming regions (below the envelope empirically defined by \citealt{2003Kauffmann_mnra346}) and AGNs (above the envelope defined by \citealt{2002Kewley_apjs142}). The color bars located on the right side run from low to high values of the O abundance, the N/O ratio, the \Ha\ equivalent widths (determined from observations and from models), the ionization parameter \logU\, and the fraction of OB stars \fOB\ (from upper-left to lower-right panels respectively). We plot here only the results concerning the models where the O/H abundance is determined from the O3N2 relation from M13 ("Neb" models, see Sec.~\ref{sec:meta_grid}).

We can see from the middle panels a general trend of the models with lower \EWa\ and lower \fOB\ to be located above the \citet{2003Kauffmann_mnra346} curve, although there are also models with low \EWa\ and low \fOB\ below this curve. We can also see that there are no photoionization models with high \EWa\ and \fOB\ above the \citet{2002Kewley_apjs142} curve. Our models reproduce the observational results obtained by \cite{2015Sanchez_aap574}: below the curve, the models with higher O abundance and lower U are located in the lower right region whereas those with a low O abundance and high U are located in the upper left corner \citep{1985Evans_apjs58, 1987Veilleux_apjs63, 2012Lopez-Sanchez_mnra426}. Notice that the middle left plot uses only observed data and then show more numerous points. The color separation in the BPT diagram is clearer when using the observed \EWa\ (left panel), while the mixing of the colors is more important for the \EWa\ from the models (right panel). This is coherent with the results from Fig.~\ref{fig:EW_comp} described in the Sec.~\ref{sec:HaEW}: the observed \EWa\ may appear to be a very good diagnostic for the ionizing population (and to derive what kind of nebular region is observed), but when using the \EWa\ from the models, the situation is really less obvious.

In the lower left panel of Fig.~\ref{fig:BPT_2} we use \logU\ as color code. It shows a clear gradient of \logU\ decreasing in the decreasing \oiii/\Hb-increasing \nii/\Ha\ direction.  In the lower right panel of Fig.~\ref{fig:BPT_2} we use \fOB\ as color code. As in Fig.~\ref{fig:EW_comp}, the two extreme values of \fOB\ are dominating the distribution and are not very well separated. We can also see some regions where the HOLMES dominate the ionizing SED well inside the star forming region, while almost no OB-stars dominated regions enter the part of the BPT diagram between the Kauffman and the Kewley curves, and not at all above the Kewley curve, in agreement with the definition of these demarcation line.

\begin{figure*}[!h]
\centering
\includegraphics[width=\hsize]{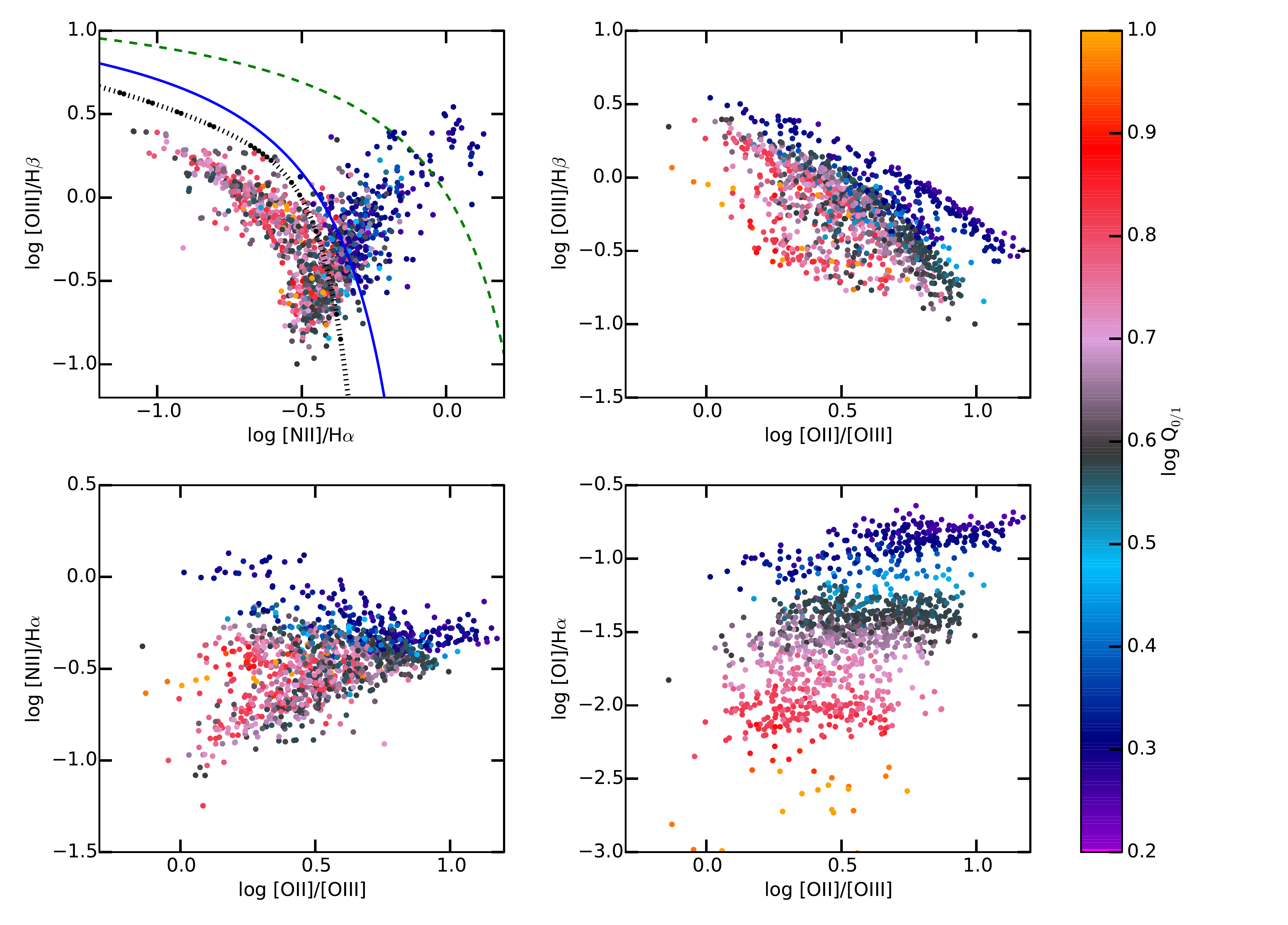}
\caption{The BPT diagrams inspired by \citet{1981Baldwin_pasp93}. The color code represents the hardness of the radiation \QHHe . The solid blue line is from \citet{2003Kauffmann_mnra346}, the green dashed line is from \citet{kewley01} and the dotted black line is from \citet{2006Stasinska_mnra371}.}
\label{fig:BPT_all}%
\end{figure*}

Figure~\ref{fig:BPT_all} shows the comparison between different BPT-type diagrams taken from \citet{1981Baldwin_pasp93, 1987Veilleux_apjs63}: $\log$(\oiii/\Hb) versus $\log$(\nii/\Ha), $\log$(\oiii/\Hb) versus $\log$(\oii/\oiii), $\log$(\nii/\Hb) versus $\log$(\oii/\oiii), and $\log$(\oi/\Ha) versus $\log$(\oii/\oiii). In these plots, we use the \QHHe\ ratio as color code, tracing the softness of the ionizing flux.

The $\log$(\oiii/\Hb) versus $\log$(\oii/\oiii) and $\log$(\oi/\Ha) versus $\log$(\oii/\oiii) ratios (upper right and lower right panels, respectively) do not depend on the N/O abundance ratio. Both plots exhibit a clear separation between the regions ionized by OB stars \QHHe\ > 0.55 and those ionized by HOLMES (otherwise), but note that in the $\log$(\oiii/\Hb) versus $\log$(\oii/\oiii) plot, some regions ionized by HOLMES are mixed with the OB star ionized regions whereas the separation in the $\log$(\oi/\Ha) versus $\log$(\oii/\oiii) plot is clearer (in agreement with the results pointed out by \citet{1981Baldwin_pasp93}. One must keep in mind that the values of \oi/\Ha\ used here are pure predictions from the models, not necessarily reproducing the observations.

\subsection{WHAN diagram}
\begin{figure*}[!h]
\centering 
\includegraphics[width=\hsize, trim = 0 0 30 0, clip = yes]{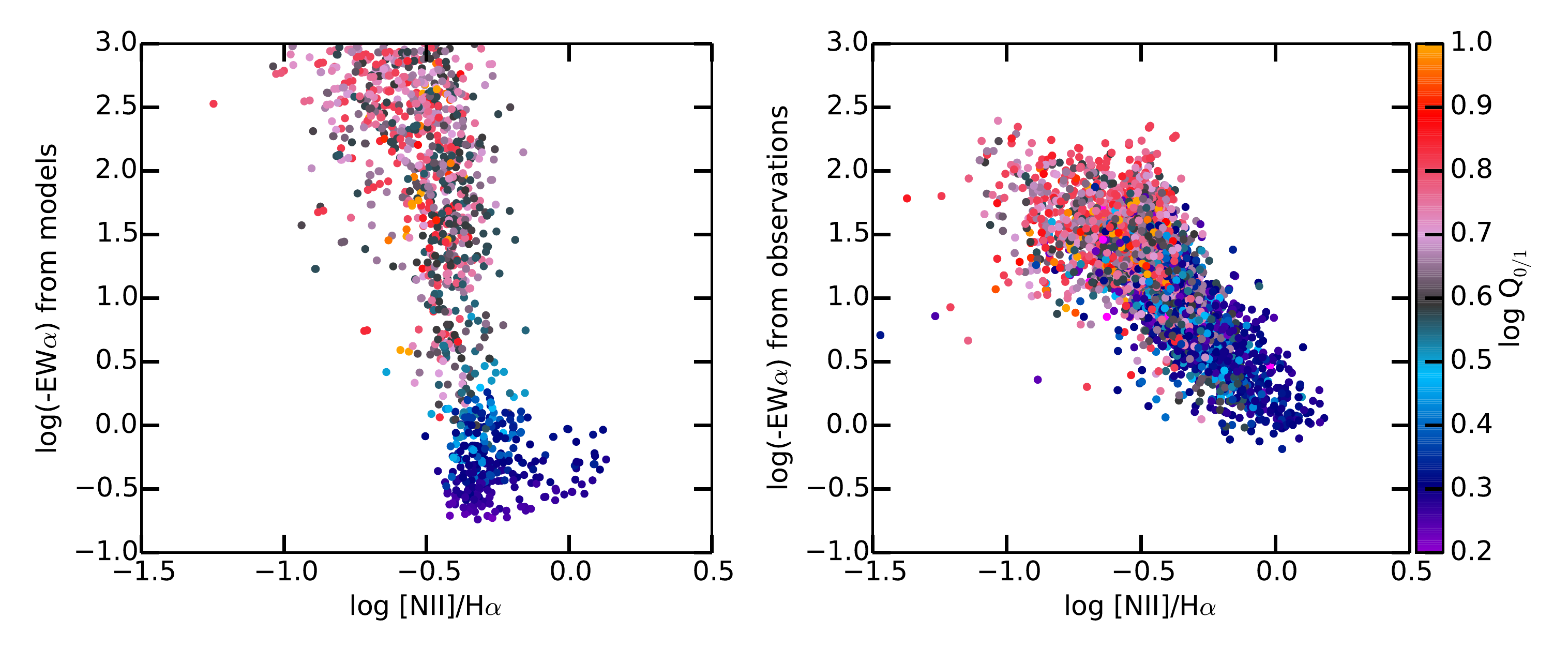}
\caption{WHAN diagrams: values of \EWa\ from the models (left panel) and from the observations (right panel) as a function of \nii/\Ha. The color code represents the proportion of OB stars.}
\label{fig:WHAN}%
\end{figure*}

Figure~\ref{fig:WHAN} shows the WHAN diagram for our models. This diagram is based on \Ha\ and \nii\ lines and was proposed by \citet{2010Cid-Fernandes_mnra403} to determine the ionizing population. In the left panel we show the \EWa\ from the models and in the right panel the ones derived from the observations. 

The general trend is that the regions ionized by OB stars have higher \EWa\ than the regions ionized by HOLMES (note that the \EWa\ from the models cover a wider range of values than the \EWa\ derived from the observations).  

The separation between the regions ionized by OB stars and those ionized by HOLMES is very clear when using the \EWa\ from the models whereas the different type of ionized regions are more mixed in the other panel. Another interesting result from this figure is that there is no correlation between the \EWa\ from the models and the \nii/\Ha\ ratios but there is a clear trend between the \EWa\ computed from the observations and the \nii/\Ha\ ratios. This indicates that if we only know the \nii/\Ha\ ratio we cannot distinguish between regions ionized by OB stars or HOLMES. Thus, the knowledge of the properties
of the underlying stellar population is a fundamental tool to distinguish between both types of regions. This was already discussed by Sanchez et al. (2014). Another interesting result is that while a limit of \EWa$_{mod}>$3\AA\ to segregate classical \HII\ regions from HOLMES is well predicted by the models, in practice, an empirical cut of \EWa > 6\AA\  or the demarcation line 
$$\log(EW_\alpha) > 2+2\log(\nii/\rm H\alpha)$$ 
seems to separate both ionizing regions better.

\subsection{\logU\ vs. \oii/\oiii\ and \sii/\siii}

\begin{figure*}[!h]
\centering
\includegraphics[width=\hsize]{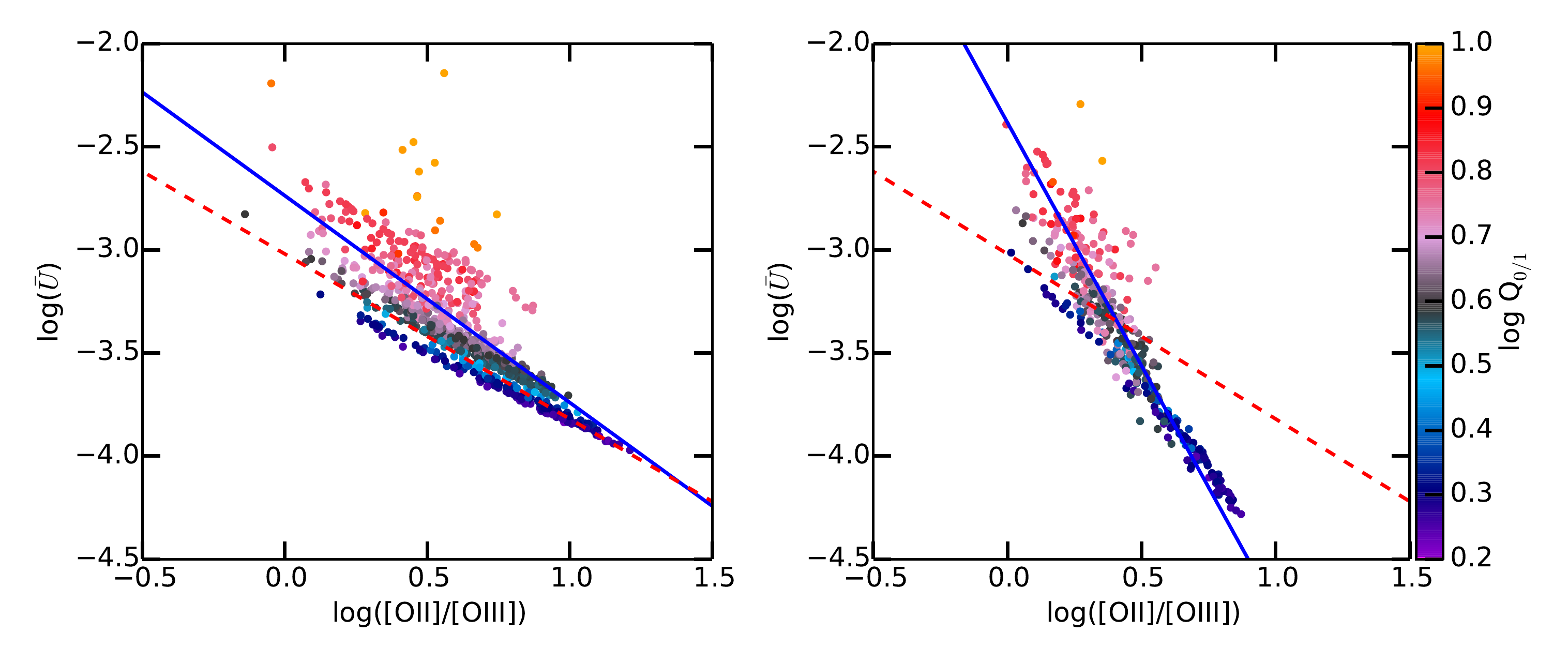}
\caption{\logU\ vs. \oii/\oiii\ for the models. The colors coding the softness of the radiation emitted by the stellar population. Left panel: thin shell models ($fr$ = 3.0), right panels: filled sphere models ($fr$ = 0.03). The dashed red line is the fit by \cite{2000Diaz_mnra318}. The blue lines correspond to our fit, taking only star forming regions into account (\QHHe\ > 0.55).
}
\label{fig:logU_O23}%
\end{figure*}

\begin{figure}[!h]
\centering
\includegraphics[width=\hsize]{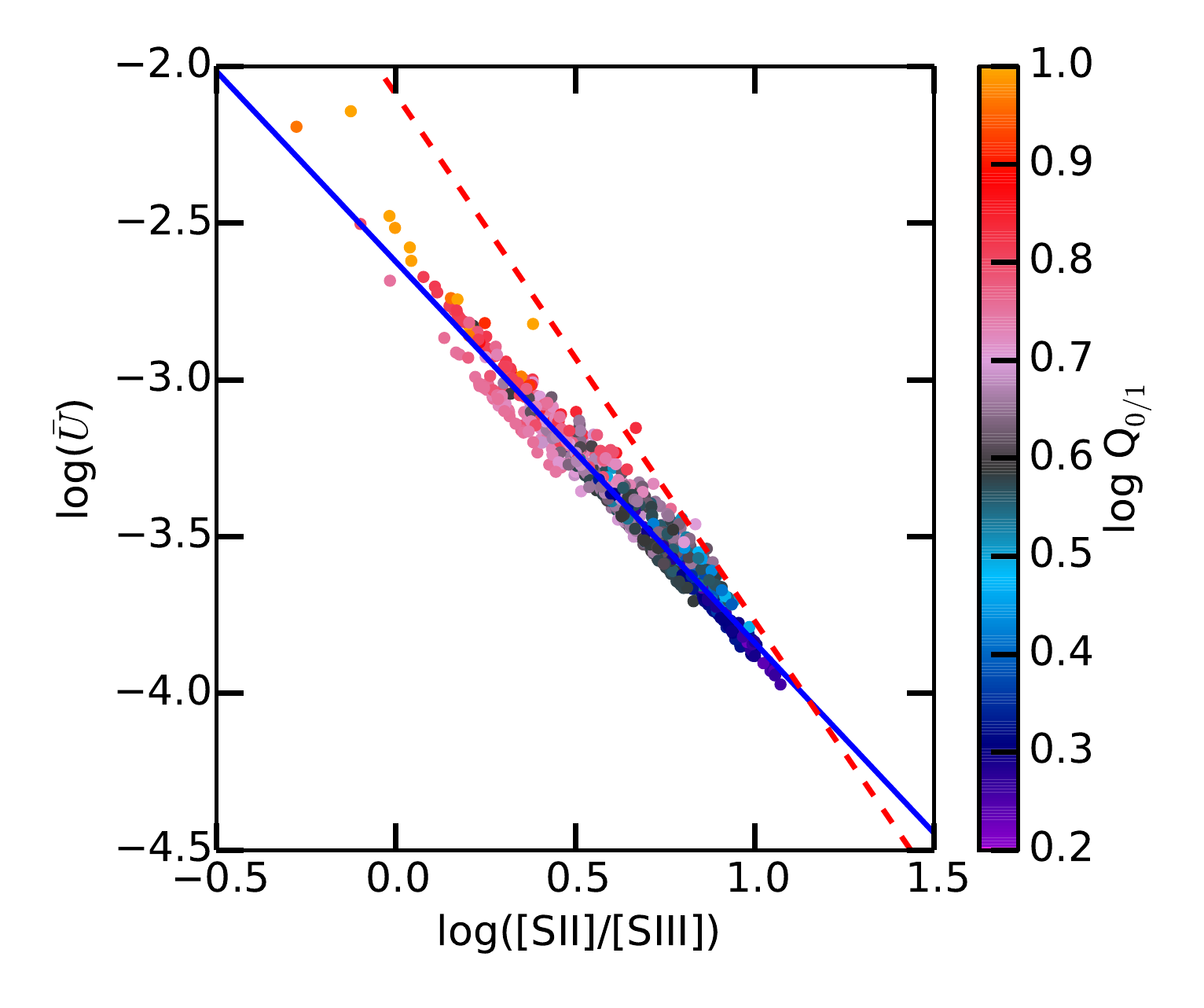}
\caption{\logU\ vs. \sii/\siii\ for the models. The colors coding the hardness of the ionizing radiation. As there is no noticeable difference due to the morphology of the region, contrary to the case of \oii/\oiii\ shown in Fig.~\ref{fig:logU_O23}, both morphologies are plotted here in the same figure. The blue line corresponds to our linear fit (taking only star forming regions into account i.e. \QHHe\ > 0.55)} see text for the values. The red dashed line corresponds to the fit by \citet{1991Diaz_mnra253}.
\label{fig:logU_S23}%
\end{figure}

The \oii/\oiii\ and \sii/\siii\ line ratios has been frequently used as tracers of the ionization strength 
\citep{2001Diaz_247, 2002Kewley_apjs142}, based on empirical correlations between these line ratios and this parameter \citep{2003Dors_aap404}. However, as we have seen along this article, previous results are sometimes based on photoinization models which ionization source was not selected to match any observed constraint. Therefore, it is important to revise the trends between those parameters based on our new set of models. In Figs.~\ref{fig:logU_O23} and \ref{fig:logU_S23}, we explore the relations between \logU\ and the \oii/\oiii\ and \sii/\siii\ line ratio, respectively. 

In Fig.~\ref{fig:logU_O23} we show the first of these relations for each of the geometries considered in our models: thin shell and empty sphere (left and right panel respectively). We overplot the relation determined by \cite[][hereafter D00]{2000Diaz_mnra318}. This relation does not match the results of our models, specially the ones corresponding to star forming regions (\QHHe\ > 0.55, gray/red points). In the case of thin shell geometry (left panel), there is an underestimation of \logU\ by $\sim$0.25 dex with a scatter of more or less the same amount. In the case of a geometrically thick region (right panel), the slope of the D00 relation is not recovered. Our results indicate that there is a steeper relation, although the average value is similar to that predicted by D00. Our fits limited to the regions where \QHHe\ > 0.55 (star forming regions) leads to: 
\begin{equation}\label{eq:logUO23a} \log(\bar U) = -2.74\pm 0.02 -1.00\pm 0.03 \times \log(\oii/\oiii)
\end{equation}
with a standard deviation of 0.14 and 
\begin{equation}\label{eq:logUO23b} \log(\bar U) = -2.38\pm 0.04 -2.36\pm 0.10 \times \log(\oii/\oiii)
\end{equation}
with a standard deviation of 0.22, for the left (thin shells) and right (filled spheres) panel respectively.

In Fig.~\ref{fig:logU_S23} we do not split the two geometries in different plots, as the results are very similar for each of them. A tighter relation is derived in comparison with the one derived for \oii/\oiii. We overplot the relation determined by \citet{1991Diaz_mnra253}, showing a different slope but similar values at low ionization. The changes in the atomic data (especially for \sii) since 1991 may explain the differences.
The following linear fit, limited to the regions where \QHHe\ > 0.55 (star forming regions), is obtained: 
\begin{equation}\label{eq:logUS23}\log(\bar U) = -2.62\pm 0.01 -1.22\pm 0.01 \times \log(\sii/\siii)
\end{equation}
with a standard deviation of 0.06. 

\subsection{N/O vs. O/H}
\label{sec:NOoOH}

The nucleosynthesis paths for Nitrogen and Oxygen are different. The origin of Nitrogen is both primary, produced from the initial content of Hydrogen, and secondary, produced from the initial content of Carbon and Oxygen created by previous stellar generations \citep[e. g.][]{1986Matteucci_mnra221}. Therefore, the N/H and O/H abundance ratios are not supposed to evolved in lockstep. The N/O vs. O/H relation derived from observations is mainly horizontal for log(O/H) < -4.1 where the Nitrogen is of primary production, and almost linear above this, where the secondary production of Nitrogen take place \citep[see e.g.][and references therein]{1993Vila-Costas_mnra265, 1996Thurston_mnra283, 2003Chiappini_mnra339, 2006Liang_apj652, 2006Molla_mnra372}. The regions we are using in this work are all above this limit and only allow us to probe the linear part of the relation.

\begin{figure*}[!h]
\centering
\includegraphics[width=\hsize]{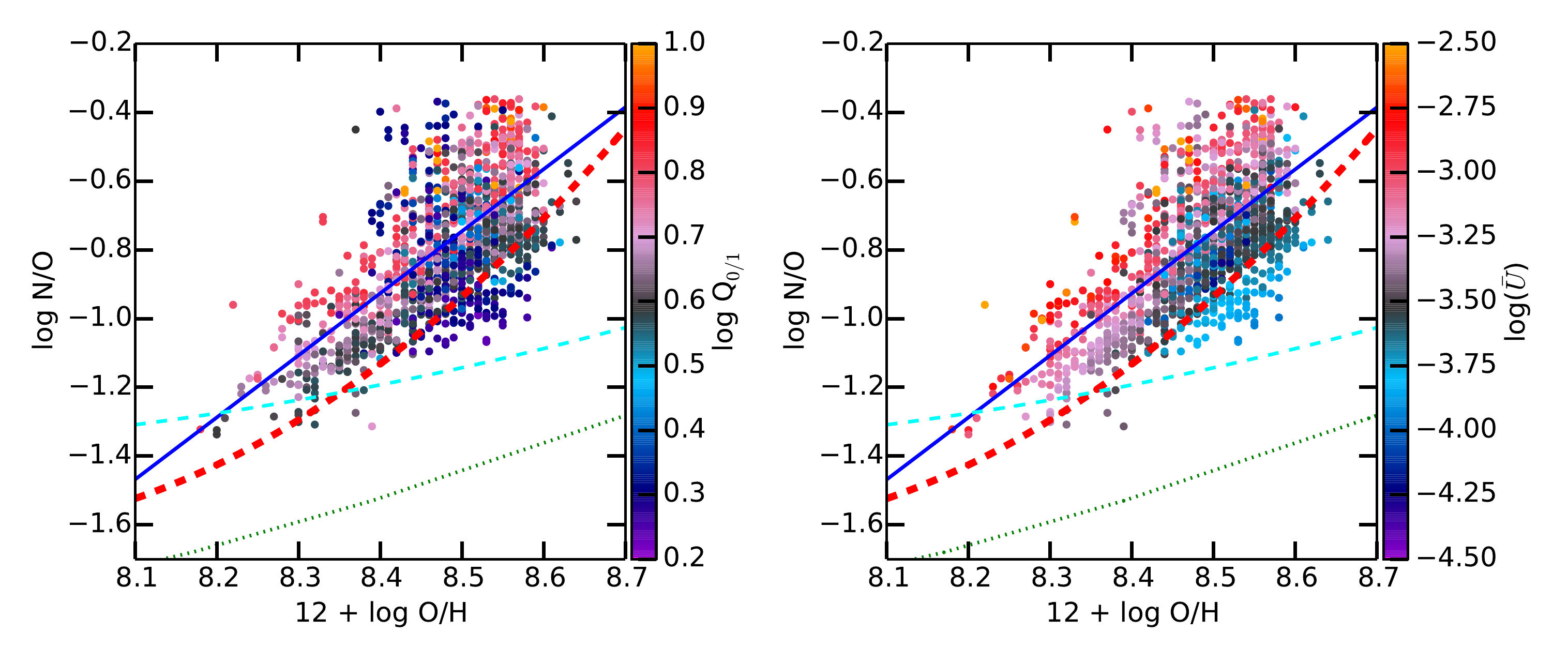}
\caption{N/O vs. O/H. Left panel: the color is coding the hardness of the ionizing radiation \QHHe. Right panel: the color is coding the value of \logU. The blue line is the fit of the \QHHe\ > 0.55 regions (gray/red/orange points), see text for the corresponding values.  The red dashed line corresponds to the fit by \citet{2012Pilyugin_mnra421}, the cyan dashed line to \citet{1993Vila-Costas_mnra265}}, and the green doted line to the fit by \citet{2013Dopita_apjs208}.
\label{fig:NO_O}%
\end{figure*}

Figure~\ref{fig:NO_O} shows the relation between the abundance ratio N/O and the oxygen abundance O/H. Left and right panels use different color code: stellar population as depicted by the softness parameter \QHHe\ and \logU\ respectively. Remember that the N/O abundance ratio is one of the main results, with \logU , of the fitting process we applied, see Sec.~\ref{sec:meta_grid}.

The difference between the regions ionized by OB stars and by HOLMES is shown in the left panel: at a given metallicity, the N/O ratio is lower in the second group (\QHHe\ < 0.55, blue points) than in the classical star forming regions (\QHHe\ > 0.55). This is also the case even if we define the nebular metallicity from the stellar content ("Stel" models, see Sec.~\ref{sec:stelabund}). These differences in the N/O determinations may be an indication that the amount of Nitrogen of the regions ionized by HOLMES is systematically lower than in the star forming regions.

If we concentrate on the star forming regions only, we can determine a linear fit to the relation between N/O and O/H: 
\begin{equation} 
\log({\rm N/O}) = -16.09\pm 0.40 + 1.81\pm 0.04 \times (12+\log({\rm O/H})).
\end{equation}
with a standard deviation of 0.13.

This fit is valid only on the O/H range available with the data used here, namely between $8.1 < 12 + \log({\rm O/H}) < 8.8$. There a clear color gradient following \logU\ in the right panel, pointing to a second order effect of the gas ionization stage on the relation.

We compare this fit to the ones determined by \citet[][light green dashed line]{2012Pilyugin_mnra421} and by \citet[][red dotted line]{2013Dopita_apjs208}. There is a good agreement with the first one, and a clear offset with respect to the second. This result is expected, since \citet{2012Pilyugin_mnra421} adopted oxygen abundances derived using a direct method, while \citet{2013Dopita_apjs208} uses a particular set of photoioniziation models that over-estimate the abundances. Again, the result confirms our previous claim that our photoinization models are compatible with direct method estimations, contrary to some previous results.

\subsection{\logU\ vs. O/H}
\label{sec:logUOH}
\begin{figure*}[!h]
\centering
\includegraphics[width=\hsize]{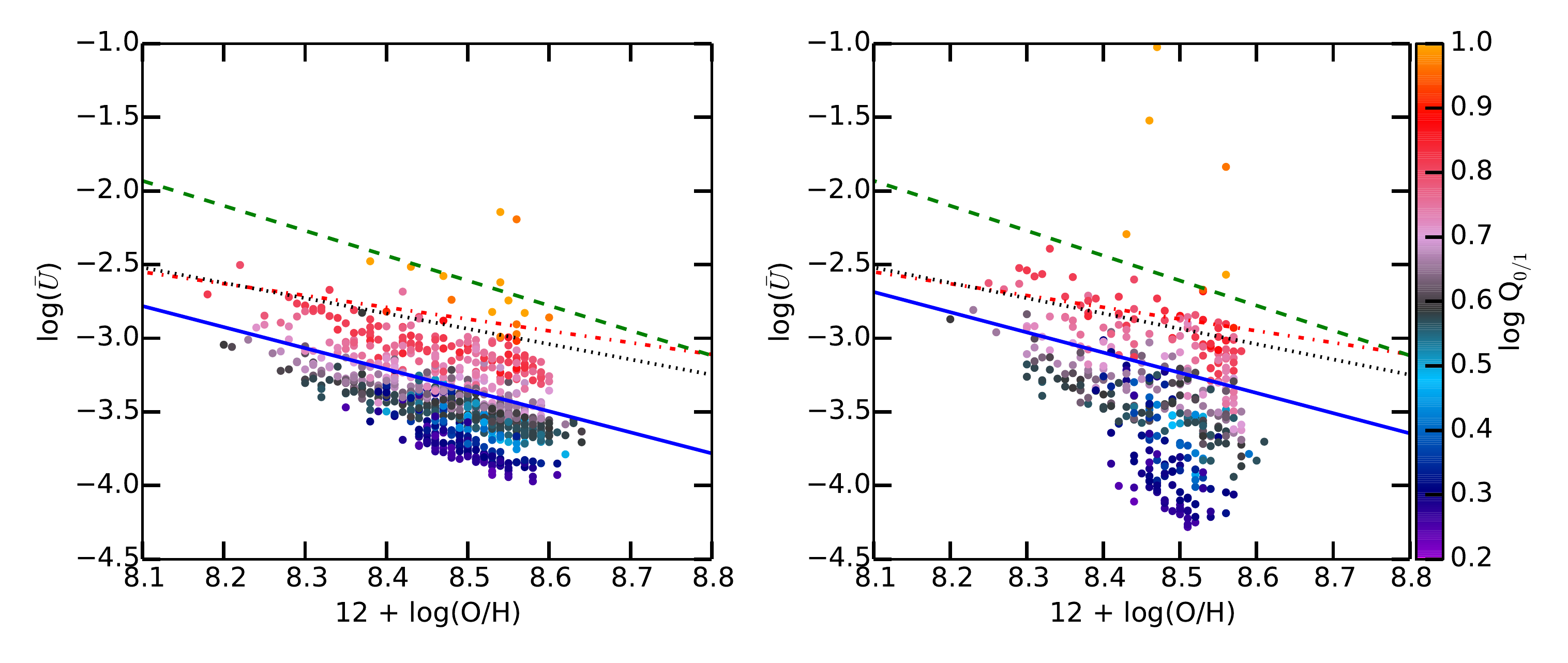}
\caption{\logU\ vs. O/H. Left panel: thin shell models ($fr$ = 3.0), right panel: filled sphere models ($fr$ = 0.03). The solid red dot-dashed line corresponds to the relation from \cite{1986Dopita_apj307}, the green dashed line to \citet{2006Dopita_apj647} and the red dotted line to a fit to the Fig. 3 from \citet{2014Perez-Montero_mnra441}. The color code is following \QHHe. Our fit is shown with the blue line, it is obtained considering only the \QHHe\ > 0.55 regions, see text for the corresponding values. }
\label{fig:logU_met}%
\end{figure*}

We plot in Fig.~\ref{fig:logU_met} the position of the models in the \logU\ vs. 12 + log(O/H) diagram. We also added the empirical relations from \citet{1986Dopita_apj307}, \citet{2006Dopita_apj647} and \citet{2014Perez-Montero_mnra441}. We can easily see that the \logU\ we determined are lower by around 0.7 dex than the values obtained from these relations. Or that our oxygen abundance is between 0.4 and 0.7 dex lower, depending on the relation we consider. This apparent discrepancy is principally due to the way we determined the oxygen abundance (using O3N2 from M13); using the abundance estimator from e. g. \citet{2002Kewley_apjs142} would lead to higher values for O/H and could reconcile the values we obtained with the different relations shown here. See also the discussion of the results obtained with the "Stel" models in Sec.~\ref{sec:stelabund}.

The effect of the ionizing stellar population (used for the color code) is also very clear. The \logU\ values for the regions ionized by HOLMES are 0.5 dex lower than for the classical HII regions (ionized by OB stars).

Our fits to the models corresponding to star forming regions (\QHHe\ > 0.55) are: 
\begin{equation}\label{eq:logUOa} \log(\bar U) =  8.79\pm 0.76 -1.43\pm 0.09 \times (12+\log({\rm O/H})) \end{equation}
with a standard deviation of 0.20 and 
\begin{equation}\label{eq:logUOb} \log(\bar U) = 8.42\pm 2.02 -1.37\pm 0.24 \times (12+\log({\rm O/H})) \end{equation}
with a standard deviation of 0.20 for the left (thin shell) and 0.36 for the right (filled sphere) panel respectively. Notice the very high uncertainties and dispersion in the case of filled sphere models.

\subsection{The variation of the $\eta$ parameter with the ionizing SED}
\label{sec:etas}

\begin{figure*}[!h]
\centering
\includegraphics[width=\hsize]{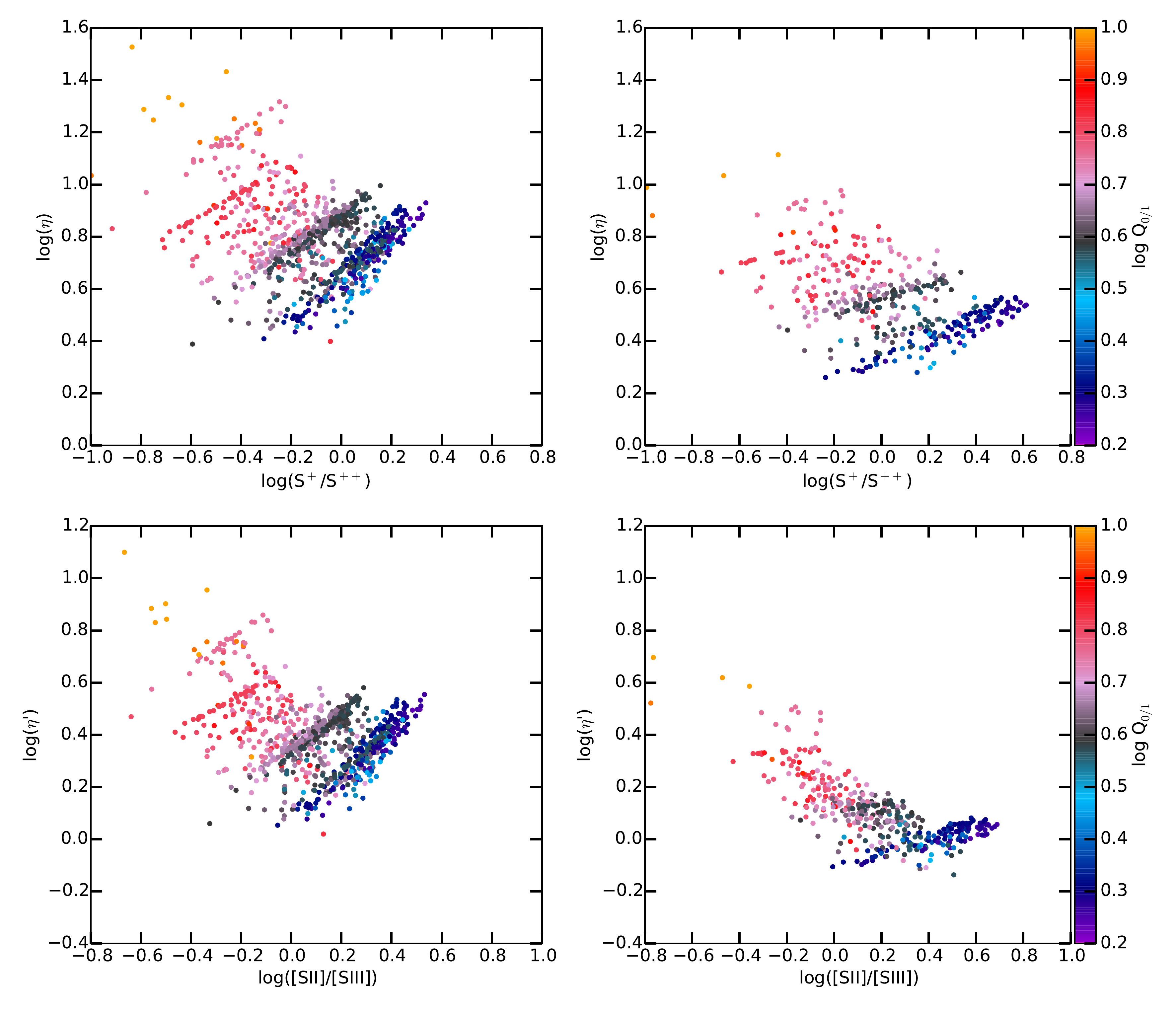}
\caption{Relation between $\eta$ = (O$^{+}$/O$^{++}$) / (S$^{+}$/S$^{++}$) and S$^{+}$/S$^{++}$ (upper panels) and between $\eta '$ = (\oii$\lambda$3727+/\oiii$\lambda$5007+) / (\sii$\lambda$6720+/\siii$\lambda$9067+)  and \sii$\lambda$6720+/\siii$\lambda$9067+ (lower panels). 
Left panel: thin shell models ($fr$ = 3.0), right panel: filled sphere models ($fr$ = 0.03). The color code follows the \QHHe\ ratio (see text).}
\label{fig:Etas}%
\end{figure*}

Following \citet{1988Vilchez_mnra231}, we plot in Fig.~\ref{fig:Etas} the values of $\eta$ = (O$^{+}$/O$^{++}$)/(S$^{+}$/S$^{++}$) vs. S$^{+}$/S$^{++}$ and $\eta '$ = (\oii$\lambda$3727+ / \oiii$\lambda$5007+) / (\sii$\lambda$6720+ / \siii$\lambda$9067+)  vs. \sii$\lambda$6720+ / \siii$\lambda$9067+. We show that these $\eta$ and $\eta '$ depend on the softness of the ionizing SED, represented here by \QHHe\ as color code, as already pointed out by \citet{1988Vilchez_mnra231}. But we also see here that the $\eta '$ depends strongly on the geometry too, left panel being the thin models ($fr$ = 3.0), and right panels the filled sphere models ($fr$ = 0.03). Given the scatter observed in each panel and the relatively high difference between the two geometries, the relation between the $\eta$'s and the softness of the ionizing radiation \QHHe\ is far from being established.

\subsection{Effect of the nebular abundance determination on the results}
\label{sec:stelabund}

We redraw all the figures presented in the previous sections, but now using the models obtained with the "Stel" determination of the nebular O/H (see Sec.~\ref{sec:meta_grid}). The corresponding figures are available in the online edition appendix \ref{append:stelabund}. 
The main differences between the two sets of models are obviously seen in plots directly involving the nebular metallicity, for example in the upper left panel of Figs.~\ref{fig:BPT_2}/\ref{fig:BPT_2_Stel} ("Neb"/"Stel" models resp.) where the color distribution is strongly affected. Notice that the position of the models in all the BPT diagrams are the same, as each model is reproducing the \nii/\Ha , \oii/\Hb , and \oiii/\Hb\ line ratio, whatever the way its nebular metallicity is obtained.

The results related to the relation between \logU\ and \oii/\oiii\ shown in Figs.~\ref{fig:logU_O23}/\ref{fig:logU_O23_Stel} are a little bit affected, but the main conclusion stay unchanged: the relation from \citet{2000Diaz_mnra318} is not recovered. When using \sii/\siii\ as in Figs.~\ref{fig:logU_S23}, there is a perfect match between the both sets of models. This leads to the conclusion that the results obtained here, in particular the fit from Eq.~\ref{eq:logUS23} (but also \ref{eq:logUO23a} and \ref{eq:logUO23b}), are very robust.

The more important differences are obtained for \ref{fig:NO_O}/\ref{fig:NO_O_Stel} and Figs.~\ref{fig:logU_met}/\ref{fig:logU_met_Stel} showing the relations between \logU\ and N/O vs. O/H respectively. Clear relations are not obtained at all when the nebular metallicity is derived from the stellar content ("Stel" models). Nevertheless, those two relations exist and have been observed using other methods. 
The lack of relation between \logU\ and N/O vs. O/H when using the "Stel" models indicates that 1) the "Stel" models are not correct regarding the nebular metallicity, and 2) that these relations obtained with the "Neb" models may be dependent on the way the nebular metallicity is obtained.

\section{Conclusions}\label{sec:concl}

We present in this paper a set of photoionization models based on the CALIFA HII regions catalog. Each model uses as ionizing SED the combination of POPSTAR stellar population models \citep{2009Molla_mnra398}, based on the analysis of the continuum spectra performed by the FIT3D program \citep{2011Sanchez_mnra410} for the corresponding region. Each model corresponds to an interpolation in the \logU\ vs. N/O parameter space to fit the observation of the \nii/\Ha\ and \oiii/\Hb\ line ratios of an individual HII region. Two different morphologies (filled or empty sphere) as well as two different ways to derive O/H (from O3N2 and from the stellar population) are explored. The fact that O/H is determined by strong line method leads to qualify these models as "hybrid" ones. We finally filter the models by selecting only the ones that also fit of the \oii/\Ha\ line ratio and excluding the ones that correspond to HOLMES ionizing spectrum and fall in the star forming region of the BPT diagram. We obtain a set of 2,558 models, each one fitting simultaneously the tree line ratios of a given HII region. 
This incomparable database allows us to explore relations between parameters, with the possibility to take into account the effect of the way the nebular metallicity is defined, or the morphology of the region.

The ionizing stellar population can be divided in two groups: classical OB-stars ionizing star forming regions, and HOLMES resulting from the evolution of old starbursts. Three quarters of the models correspond to star forming regions. We found that the first regions show a difference in the \Ha\ equivalent width between models and observations that can be interpreted as the result of a median leaking of 80\% of the photons. On the contrary, for the HOLMES ionized regions we found that the models predict higher values for \EWa\ than what is actually observed. This can be understood as missing ionizing photons compared to what would be needed to produce the HI recombination lines.

We show that our models are mainly compatible with the electron temperature derived from observations for a given value of O/H, which was not the case for previously publish grids of models \citep[][e.g.]{2013Dopita_apjs208}. We attribute the better match of our models to the use of a detailed ionizing SED obtained from the stellar underlying population for each region.

We derive new relations between \logU\ vs. \oii/\oiii\ and \sii/\siii, showing that the first one strongly depends on the morphology of the nebula, while the latest one is a very robust result (it does not depend on the way the nebular abundance is determined).

The relation between N/O and O/H we derive is compatible with \citet[][using a method based on observations]{2012Pilyugin_mnra421} and not with \citet[][based on photoionization models]{2013Dopita_apjs208}. 
The relation between \logU\ and O/H is different from the previous determinations, leading for lower values at a given metallicity. We also conclude that $\eta '$ is not a good indicator of the softness of the radiation field, as it also strongly depends on the morphology of the region.

All the figures presented in this paper can easily be generated by anyone given that the data are available from 3MdB and that the python codes used to make the models and the figures are available from the 3MdB web page \url{https://sites.google.com/site/mexicanmillionmodels}.

\begin{acknowledgements}
CM took advantages of the useful discussions with G. Hägele and his co-workers while invited at La Plata University.
\\
We thank the anonymous referee for her/his careful review of the paper and valuable comments.
\\
We want to thank Yago Ascasibar and Polychronis Papaderos, the internal CALIFA referee for this paper, for their useful comments.
\\
Part of the results presented here have been obtained using computers from the CONACyT project CB2010/153985 and the UNAM-PAPIIT project 107215. The computers "Tychos" (Posgrado en Astrofísica-UNAM, Instituto de Astronomia-UNAM and PNPC-CONACyT) have been used for this research.
\\
GDI gratefully acknowledges support from the Mexican CONACYT grant CB-2014-241732.
\\
SFS thanks the CONACYT-125180 and DGAPA-IA100815 projects for providing him support in this study.
\\
Support for LG is provided by the Ministry of Economy, Development, and Tourism's Millennium Science Initiative through grant IC120009, awarded to The Millennium Institute of Astrophysics, MAS. LG acknowledges support by CONICYT through FONDECYT grant 3140566.
\\
This study uses data provided by the Calar Alto Legacy Integral Field Area (CALIFA) survey
(\url{http://califa.caha.es/}).
CALIFA is the first legacy survey performed at Calar Alto. The CALIFA collaboration would like to thank the IAA-CSIC and MPIA-MPG as major partners of the observatory, and CAHA itself, for the unique access to telescope time and support in manpower and infrastructures. The CALIFA collaboration also thanks the CAHA staff for the dedication to this project.
\\
Based on observations collected at the Centro Astronómico Hispano Alemán (CAHA) at Calar Alto, operated
jointly by the Max-Planck-Institut für Astronomie and the Instituto de Astrofísica de Andalucia (CSIC).

\end{acknowledgements}
%-------------------------------------------------------------------
\bibliographystyle{aa}
\bibliography{BIBLIO_min,CALIFAI}

\Online

\begin{appendix} 
\section{Results obtained from the "Stel" models}
\label{append:stelabund}

In this on-line appendix, we present the same figures as already shown in the paper, but obtained with the "Stel" abundance determination, instead of the "Neb" one (See Secs.~\ref{sec:meta_grid} and \ref{sec:stelabund}). We only show the figures significantly different from the "Neb" case.

\begin{figure*}[!h]
\centering
\includegraphics[width=\hsize]{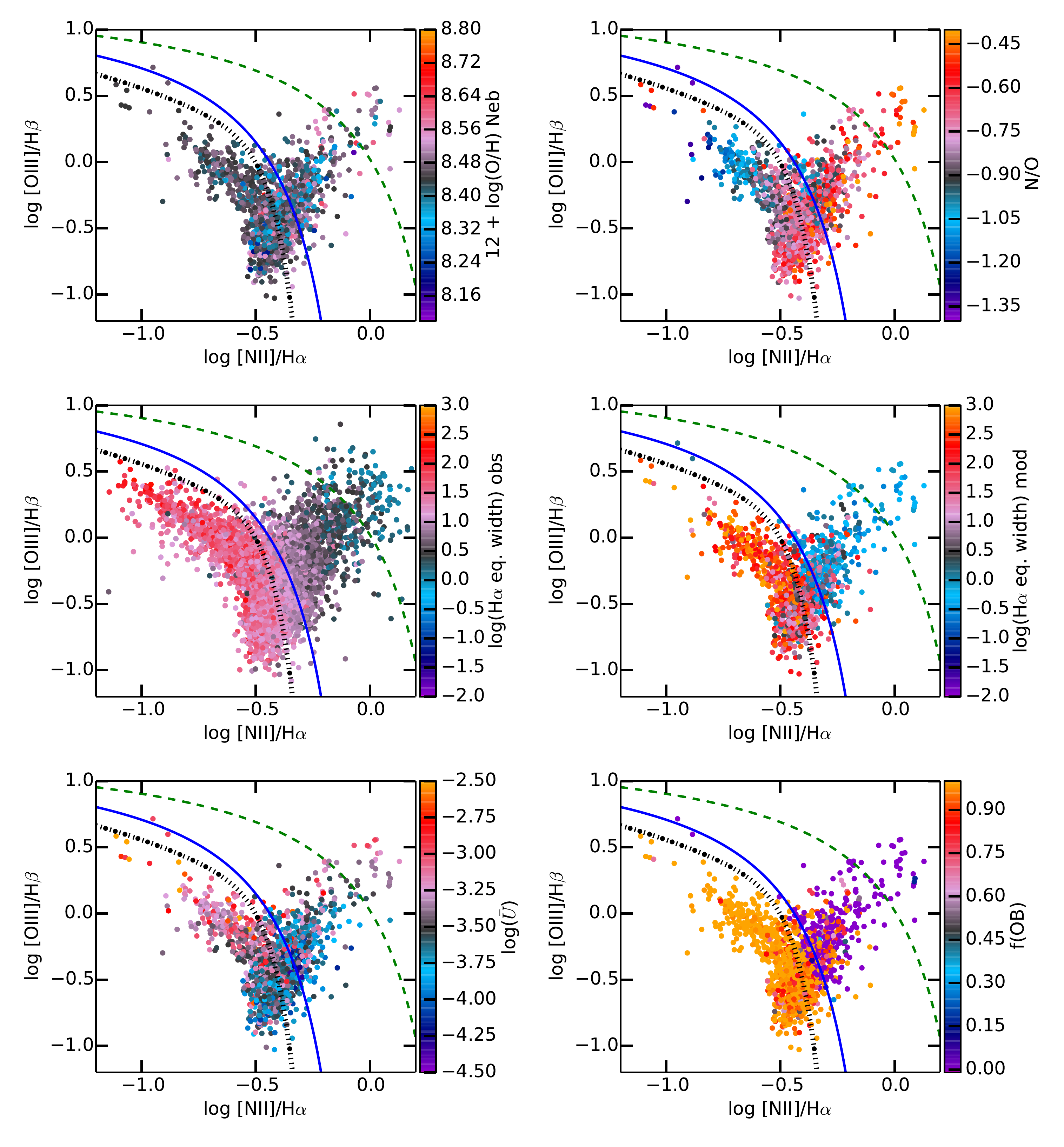}
\caption{Same as Fig.~\ref{fig:BPT_2} but using the "Stel" models.}
\label{fig:BPT_2_Stel}%
\end{figure*}

\begin{figure*}[!h]
\centering
\includegraphics[width=\hsize]{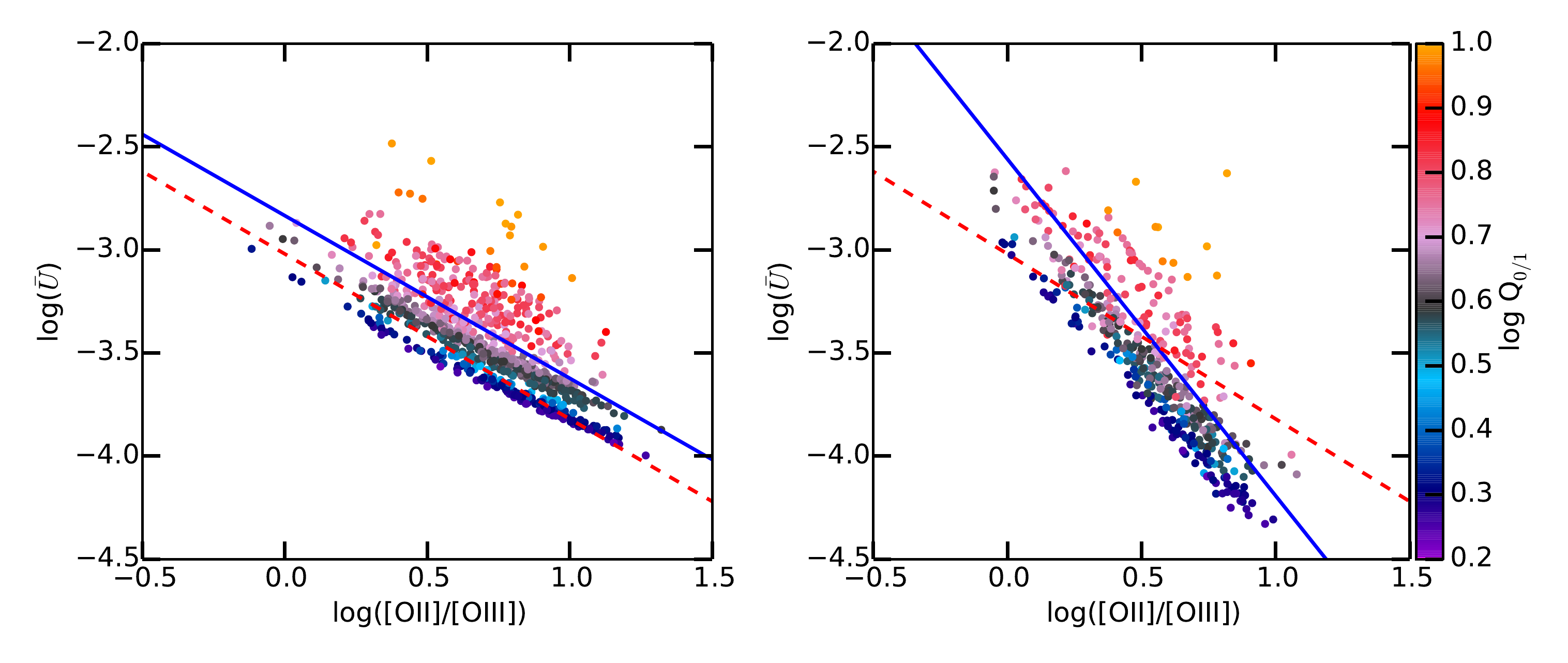}
\caption{Same as Fig.~\ref{fig:logU_O23} but using the "Stel" models.}
\label{fig:logU_O23_Stel}%
\end{figure*}

\begin{figure*}[!h]
\centering
\includegraphics[width=\hsize]{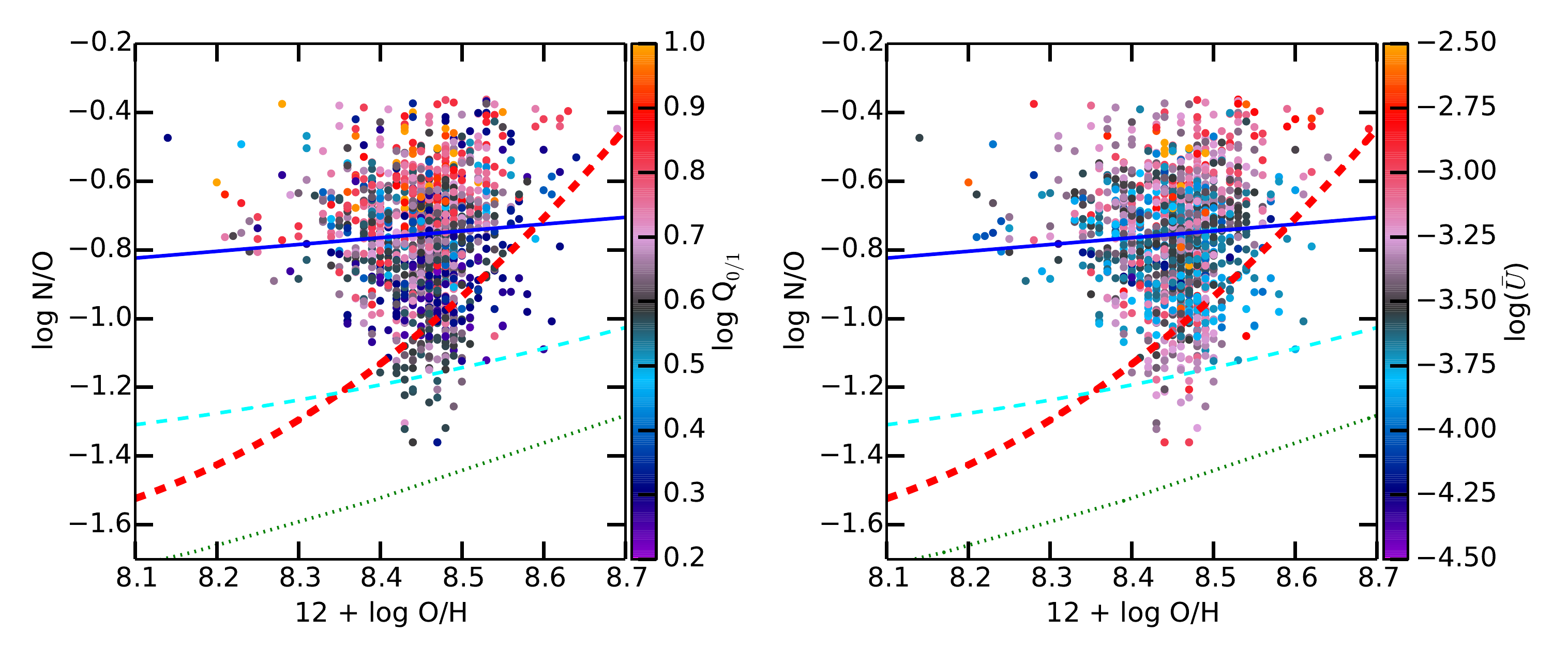}
\caption{Same as Fig.~\ref{fig:NO_O} but using the "Stel" models.}
\label{fig:NO_O_Stel}%
\end{figure*}

\begin{figure*}[!h]
\centering
\includegraphics[width=\hsize]{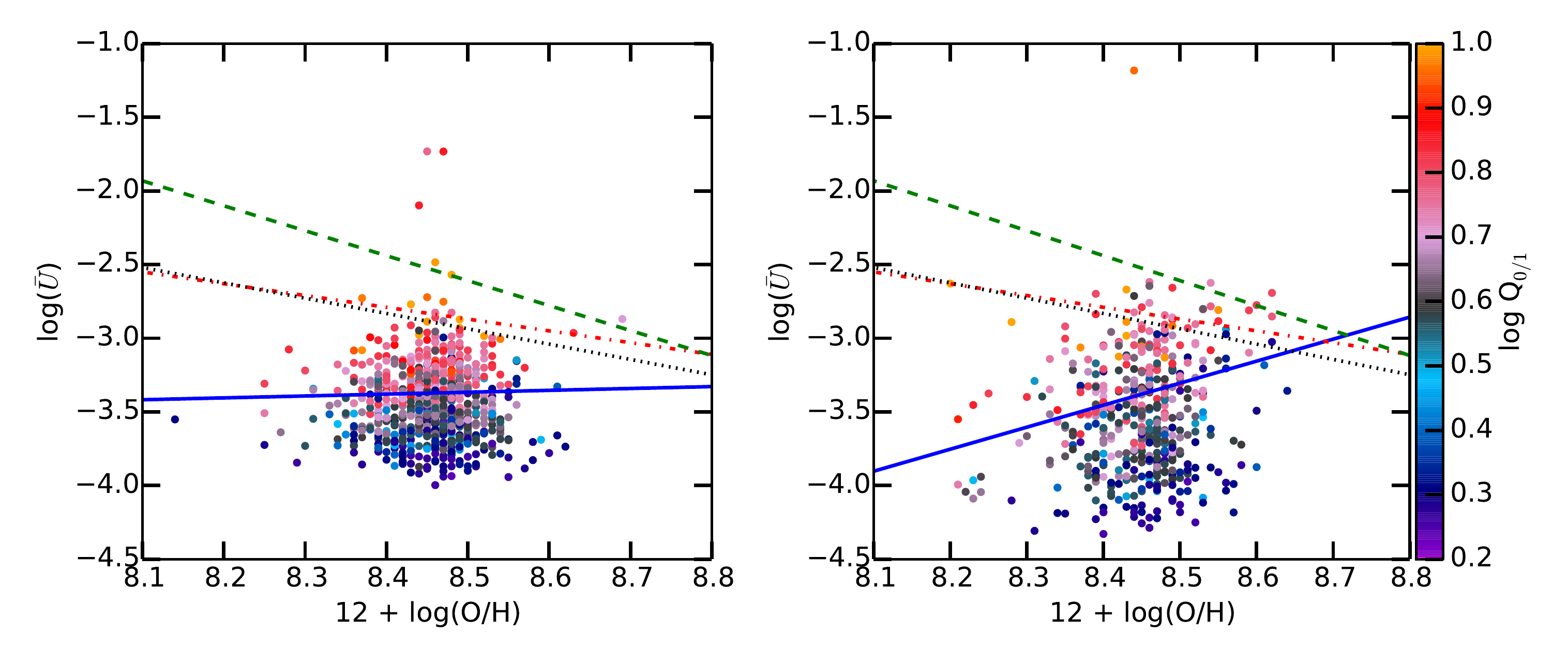}
\caption{Same as Fig.~\ref{fig:logU_met} but using the "Stel" models.}
\label{fig:logU_met_Stel}%
\end{figure*}

\end{appendix}

\end{document}